\documentclass[12pt,preprint]{aastex}

\newcommand{\lta}{\lower 0.5ex\hbox{$\buildrel < \over \sim\ $}}

\shorttitle{Spectroscopic Detection of G-modes in PG\,1627+017}
\shortauthors{For et al.}

\begin{document}

\title{First Attempt at Spectroscopic Detection of Gravity Modes in a
Long-Period Pulsating Subdwarf B Star -- PG\,1627+017\altaffilmark{1}} 


\author{B.-Q.\ For\altaffilmark{2}, E.M.\ Green\altaffilmark{2},
D.\ O'Donoghue\altaffilmark{3}, L.L.\ Kiss\altaffilmark{4}, 
S.K.\ Randall\altaffilmark{5}, \\ G.\ Fontaine\altaffilmark{5},
A.P.\ Jacob\altaffilmark{4}, S.J.\ O'Toole\altaffilmark{6}, 
E.A.\ Hyde\altaffilmark{2}, T.R.\ Bedding\altaffilmark{4}}

\altaffiltext{1}{Observations reported here were obtained at the MMT Observatory,
a joint facility of the Univerity of Arizona and the Smithsonian Institution.}

\altaffiltext{2}{Steward Observatory, University of Arizona, Tucson, 
AZ 85721; bfor@as.arizona.edu; egreen@as.arizona.edu; elainahyde@yahoo.com}

\altaffiltext{3}{South African Astronomical Observatory, P.O. Box 9, Observatory, 
7935, South Africa; dod@saao.ac.za}

\altaffiltext{4}{School of Physics, University of Sydney, NSW 2006,
Australia; laszlo@physics.usyd.edu.au; ande@physics.usyd.edu.au;
bedding@physics.usyd.edu.au}

\altaffiltext{5}{D\'epartement de Physique, Universit\'e de Montr\'eal, 
C.P.6128, Succ.\,Centre-Ville, Montr\'eal, Qu\'ebec, Canada H3C3J7; 
randall@astro.umontreal.ca; fontaine@astro.umontreal.ca}

\altaffiltext{6}{Dr.Remeis-Sternwarte, Astronomisches Institut der Universit\"at  
Erlangen-N\"urnberg, Sternwartstr. 7, 96049 Bamberg, Germany; 
otoole@sternwarte.uni-erlangen.de}


\begin{abstract}

In the first spectroscopic campaign for a PG\,1716 variable (or
long-period pulsating subdwarf B star), we succeeded in detecting
velocity variations due to g-mode pulsations at a level of
1.0--1.5~km~s$^{-1}$, just above our detection limit.  The
observations were obtained during 40 nights on 2~m class telescopes in
Arizona, South Africa, and Australia.  The target, PG\,1627+017, is
one of the brightest (V = 12.9) and largest amplitude ($\sim$0.03 mag)
stars in its class.  It is also the visible component of a post-common
envelope binary.  Our final radial velocity data set includes 84 hours
of time-series spectroscopy over a time baseline of 53 days, with
typical errors of 5--6~km~s$^{-1}$ per spectrum.  We combined the
velocities with previously existing data to derive improved orbital
parameters.  Unexpectedly, the velocity power spectrum clearly shows
an additional component at twice the orbital frequency of
PG\,1627+017, supporting Edelmann et al.'s recent results for several
other short-period subdwarf B stars, which they claim to be evidence
for slightly elliptical orbits.  Our derived radial velocity amplitude
spectrum, after subtracting the orbital motion, shows three potential
pulsational modes 3--4$\sigma$ above the mean noise level of
0.365~km~s$^{-1}$, at 7201.0~s (138.87~$\mu$Hz), 7014.6~s (142.56
$\mu$Hz) and 7037.3~s (142.10~$\mu$Hz).  Only one of the features is
statistically likely to be real, but all three are tantalizingly close
to, or a one day alias of, the three strongest periodicities found in
the concurrent photometric campaign.  We further attempted to detect
pulsational variations in the Balmer line amplitudes.  The single
detected periodicity of 7209~s, although weak, is consistent with
theoretical expectations as a function of wavelength.  Furthermore, it
allows us to rule out a degree index of $l$ = 3 or $l$ = 5 for that
mode.  Given the extreme weakness of g-mode pulsations in these stars,
we conclude that anything beyond simply detecting their presence will
require larger telescopes, higher efficiency spectral monitoring over
longer time baselines, improved longitude coverage, and increased
radial velocity precision.

\end{abstract}

\keywords{stars: extreme horizontal branch -- stars: interiors 
-- oscillations -- subdwarfs -- stars: individual (\objectname{PG\,1627+017})}

\section{Introduction}

Subdwarf B (sdB) stars are evolved hot compact stars (22,000~K $<$
T$_{\rm eff} <$ 40,000K, 5.0 $<$ log~{\it g} $<$ 6.2) that are
commonly found in the disk of our Galaxy (Saffer et al.\ 1994). They
are core helium-burning stars with extremely thin hydrogen envelopes
($<$~0.01$M_{\sun}$), believed to have masses of about 0.5$M_{\sun}$
(Saffer et al.\ 1994; Heber 1986).  There is considerable interest in
their evolutionary history, in part because a large fraction of sdB
stars occur in post-common envelope binaries ({\it e.g.}\ Allard et
al.\ 1994; Green, Liebert \& Saffer 1997; Maxted et al.\ 2001,
Morales-Rueda et al.\ 2003).  Their structural details provide an
important independent test of stellar evolution theory because the
vast majority of low mass stars are expected to develop nearly
identical helium cores during their red giant phase.  Within the last
decade, two different types of multimode pulsators have been found
among sdB stars: shorter period variables exhibiting mainly pressure
(p-)modes, and longer period variables whose pulsations must be due to
gravity (g-)modes.  These discoveries have opened up the possibility
of using asteroseismology to investigate the structure of sdB stars,
and, by extension, the helium-burning cores of most red giants.

The first short-period pulsating sdB stars were discovered by
astronomers from the South African Astronomical Observatory (Kilkenny
et al.\ 1997), and are commonly called EC\,14026 stars\footnote{Now
formally named the V361 Hya stars}, after the prototype.
Independently and nearly simultaneously, their existence was predicted
by Charpinet et al.\ (1996, 1997), based on a $\kappa$-mechanism
associated with the radiative levitation of iron in the thin
diffusion-dominated envelopes.  The resulting low-order and low-degree
($l$=0,1,2,3) radial and nonradial p-modes produce velocity variations
primarily along the star's radial axis.  Typical EC\,14026 stars have
pulsation periods of 100--200~s with amplitudes of a few hundredths of
a magnitude or less.  As predicted by Charpinet et al., they are found
mostly among the hotter sdB stars, clustering around T$_{\rm eff}$
$\sim$ 33,500~K and log~{\it g} $\sim$ 5.8, although a few cooler,
lower gravity examples exist with somewhat longer periods (up to about
10 minutes).  The current tally of published EC\,14026 stars is 34
(Kilkenny 2002a, and references therein; Silvotti et al.\ 2002;
Dreizler et al.\ 2002; Bonanno et al.\ 2003; Oreiro et al.\ 2004;
Solheim et al.\ 2004; Baran et al.\ 2005), although more than half are 
relatively faint (V~$<$~14).

The second class of sdB pulsators was serendipitously discovered by
Green et al.\ (2003) during a light curve survey originally intended
to search for binary eclipses and reflection effects.  PG\,1716 stars
(Reed et al.\ 2004), sometimes called long-period sdB variables,
pulsate on timescales of approximately an hour.  They have extremely
small photometric amplitudes, of the order of millimags (Randall et
al.\ 2004, 2005bc).  In contrast to the EC\,14026 stars, long-period
sdB pulsators are found only among cooler sdB stars, clustering
around T$_{\rm eff}$ $\sim$ 27,000~K and log~{\it g} $\sim$ 5.4.
Another difference is that the slower pulsators are much more common.
Thirty of these variables, roughly 80\% of sdB stars with temperatures
between 24000~K and 29500~K, have been confirmed since 2001, all but
two brighter than V~=~14 (Green et al., in preparation).  Fontaine et
al.\ (2003) showed that PG\,1716 stars can be excited by the same
$\kappa$-mechanism proposed by Charpinet et al., if their g-modes are
due to higher-degree ($l$=3,4,5...)  nonradial pulsations.  Thus,
although the analogy is not perfect, sdB p-mode and g-mode pulsators
bear a strong resemblance to the main sequence $\beta$ Cephei and
Slowly Pulsating B stars, respectively.

Pulsational modes are most easily detected photometrically.  Since the
first discovery, several groups have conducted surveys to search for
new EC\,14026 stars (Kilkenny 2002b, and references therein;
Bill\`eres et al.\ 2002; Solheim et al.\ 2004) and PG\,1716 stars
(ongoing Steward Observatory survey).  There have now been a number of
photometric campaigns on selected short-period pulsators (e.g.,
Kilkenny et al.\ 2002b), and more recently, long-period pulsators
(Reed et al.\ 2004; Randall et al.\ 2004, 2005bc).  The common goal is
to detect a sufficient number of pulsational modes for
asteroseismology.  Unfortunately, experience has shown that photometry
alone, especially if carried out on small telescopes, is usually
insufficient to identify these modes unambiguously with a single
theoretical model.  Unique identifications have been achieved for less
than a handful of sdB pulsators (all observed on a 4 m class
telescope), most notably PG\,0014+067 (Brassard et al.\ 2001),
PG\,1219+534 (Charpinet et al.\ 2005b), and Feige\,48 (Charpinet et
al.\ 2005a), but also including PG\,1047+003 (Charpinet, Fontaine, \&
Brassard 2003).

Charpinet et al.\ (2000) pointed out that spectroscopic detection of
radial velocity (RV) modes should contain complementary information
that might help to solve the mode identification problem.  The
brightest EC\,14026 star, PG\,1605+072, was the ideal first target from
an observational point of view, since it has the largest photometric
amplitude variations (0.2 mag) of any sdB pulsator, and a very rich
pulsational spectrum (Kilkenny et al.\ 1999). Due to its unusually low
surface gravity, it also has the longest periods ($\sim$200--550s) of any
EC\,14026 variable.  O'Toole et al.\ (2000) were the first to detect
radial velocity shifts in PG\,1605+072 using medium-resolution
time-resolved spectroscopy.  Additional time-series spectroscopy has
since been carried out at the 4~m Anglo Australian Telescope and the
William Herschel Telescope (Woolf, Jeffery, \& Pollacco 2002a) at higher
spectral ($<$ 1\AA) and temporal resolution, and also at the Danish
1.5~m and the Mt.\ Stromlo 1.9~m by O'Toole et al.\ (2002).
Woolf et al.\ resolved three pulsation modes, while O'Toole et al.\
identified five modes, all of which were present in Kilkenny et al.'s
(1999) photometry.  In 2002, an ambitious world-wide campaign, the
Multi-Site Spectroscopic Telescope (MSST), was organized to resolve
closely spaced periods in PG\,1605+072 and to obtain more precise
velocity amplitudes for weaker modes (Heber at al.\ 2003; O'Toole et
al.\ 2004).  This resulted in 151 hours of spectroscopy, leading to at
least 20 detected periodicities in PG\,1605+072.

Woolf, Jeffery, \& Pollacco (2002b) also investigated KPD\,1930+2752,
a fainter EC\,14026 variable (V~=~13.8), but found no significant
peaks in the RV power spectrum above their $\sim$~4~km~s$^{-1}$
detection limit that matched known photometric modes.  Telting \& 
\O stensen (2004) succeeded in recovering the one dominant 138~s
pulsation mode in PG\,1325+101 (also V~=~13.8) at an amplitude of
nearly 12~km~s$^{-1}$, using spectra from three nights on the Nordic
Optical Telescope.  However, it is apparent that it will be quite
difficult to obtain high quality spectroscopic detections of a
sufficiently large number of frequencies to clarify mode
identification in significantly fainter, shorter period, and lower
amplitude p-mode sdB pulsators, given the short exposures required and
the difficulties of obtaining large blocks of time on bigger
telescopes.

During the planning stages of the first multisite photometric campaign
on a long-period sdB pulsator, PG\,1627+017 (Randall et al.\ 2004),
which took place during April--June 2003, it was natural to consider
the feasibility of a complementary spectroscopic campaign to detect
g-mode pulsations.  Near-simultaneous spectroscopy and photometry are
essential, because the amplitude of individual pulsation modes can
change significantly over time (e.g.\ Kilkenny et al.\ 1999; O'Toole
et al.\ 2002).  Furthermore, assessing the reality of marginal
spectroscopic detections is much simpler if the same frequencies are
also present in the photometry.

One advantage of spectroscopy of long-period PG\,1716
variables is that they are more common than their short-period
EC\,14026 counterparts, and thus there exist more bright targets for
future investigations if spectroscopically observed g-modes prove
helpful for mode identification.  A primary disadvantage is that
their pulsational amplitudes are significantly smaller.  The fact that
typical PG\,1716 pulsational periods are roughly a factor of thirty
longer than the periods in EC\,14026 stars is both a blessing and a
curse.  Periods of an hour or longer can be more easily sampled
spectroscopically using smaller telescopes, but the effective time
baseline and the number of telescope nights required for comparable
frequency resolution must also be about thirty times larger.  From the
ground, multisite observations to reduce daily aliases are much more
critical for the long-period sdB variables, since only 4--10
cycles can be observed per night from any one site.

With these factors in mind, we organized a spectroscopic collaboration
between Steward Observatory (SO) in Arizona, the South African
Astronomical Observatory (SAAO), and Siding Spring Observatory (SSO)
in New South Wales, Australia, which would potentially allow nearly
24~hour coverage of PG\,1627+017 with 2~m class telescopes.  Ideally,
we hoped to get as much overlapping time as possible during an
approximately 7--10 day period in May, with additional nights
at two and four week intervals before and after the joint time in
May, to improve the time baseline.  As will be seen in \S 4, the
telescope allocation committees were as generous as we could have
wished, but the weather was considerably less so.  In the end, we were
able to acquire (barely) sufficient data for a productive first look
at g-modes in long-period sdB stars.
 
One additional point deserves consideration.  In addition to RV
variations, the relative pulsational amplitudes as a function of
wavelength provide another way to constrain mode identifications.
Theory predicts that pulsational amplitudes should be much greater at
the central wavelengths of the Balmer lines than in the nearby
continuum.  O'Toole et al. (2003) found a significant increase in
their line index flux for higher order Balmer lines in PG\,1605+072.
The continuum amplitudes also increase slowly from redder wavelengths
to bluer ones, with a particularly rapid increase from blue to
ultraviolet.  The ratio of UV/red flux is a strong function of the
degree index {\it l} (Randall et al.\ 2004, 2005a).  To help investigate
the amplitude dependence on wavelength in our PG\,1627+017 data, we
obtained a number of additional spectra for the similar but apparently
nonvariable sdB, PG\,1432+004, in order to calculate the atmospheric
extinction correction with wavelength.

In the remainder of this paper, we describe our multisite time-series
spectroscopy for PG\,1627+017 and present the results. In \S 2 and \S 3, 
we discuss the observed characteristics of our target and the
feasibility of detecting g-modes in this star.  \S 4 and \S 5 
describe the observations obtained at each site and the reduction methods.  
A description of the radial velocity cross-correlations and the derived
orbital parameters is given in \S 6.  We discuss the power spectrum
of the residual velocities in \S 7, and consider the wavelength
dependence of the pulsational amplitudes in \S 8.   \S 9 contains the 
summary and conclusions.

\section{PG\,1627+017}

PG\,1627+017 (16$^h$29$^m$35$^s$.3,
+01$^{\degr}$38$^{\arcmin}$19$^{\arcsec}$) is one of the brightest
known long-period sdB pulsators.  It is easily observed from both
hemispheres, and as shown in Figure~1, the $\sim$0.03 mag amplitude
variations are relatively large for its type.  At V = 12.899
$\pm$~0.019 (Allard et al.\ 1994), it has essentially the same
brightness as PG\,1605+072, y = 12.914 $\pm$~0.072 (Wesemael et al.\
1992).

PG\,1627+017 is also one of the coolest sdB stars, and has one of the
lowest surface gravities.  Morales-Rueda et al.\ (2003) derived
T$_{\rm eff}$ = 21600, log~{\it g} = 5.12, and log~(N(He)/N(H)) =
$-$2.9 using solar metallicity line-blanketed LTE atmospheres.  We
have obtained additional time-averaged, high signal-to-noise spectra
as part of a continuing program to provide homogeneous atmospheric
parameters of a large sample of sdB stars.  We use the MMT Blue
spectrograph to acquire medium resolution (1\AA) spectra, and the
Steward 2.3~m spectrograph for low resolution (8.7\AA) spectra.  We
fit our two PG\,1627+017 spectra to hydrogen and helium line blanketed
NLTE atmospheres (Green, Fontaine, \& Chayer, in preparation) to
obtain 23669 $\pm$ 190~K, log~{\it g} = 5.315 $\pm$~0.021,
log~(N(He)/N(H)) = $-$2.968 $\pm$~0.029 from the MMT spectra, and
23987 $\pm$~257~K, log~{\it g} = 5.250 $\pm$~0.034, log~(N(He)/N(H)) =
$-$2.901 $\pm$~0.105 from the Steward spectra.\footnote{The
significant difference in our effective temperatures compared to
Morales-Rueda et al.'s is due to the use of models with zero metals vs
solar metallicity, neither of which adequately describes the
diffusion-dominated outer layers of sdB stars.  More realistic sdB
temperatures will only be obtained using customized model atmospheres
appropriate for each individual metallicity distribution, however,
PG\,1627+017's relative position in the log~{\it g} vs. T$_{\rm eff}$
diagram remains the same regardless of which models are used.}  The
cool temperature of PG\,1627+017 is consistent with the longest
observed quasi-periods ($\sim$2$^h$) yet found in any of the known
long-period sdB pulsators, while the low gravity contributes to the
relatively large pulsational amplitudes seen in Figure~1.  Details of
the photometric pulsation periods and amplitudes can be found in
Randall et al.\ (2004; 2005b).

Finally, we note that PG\,1627+017, like many sdB stars, is a
post-common envelope binary.  Morales-Rueda et al.\ (2003) 
derived an orbital period of 0.829226 day and an amplitude of
73.5~km~s$^{-1}$.

\section{Feasibility}

Radial velocity precision is obviously a crucial consideration for
successful spectroscopic detection of the smaller g-mode amplitudes.
We were encouraged by our previous success during the MSST campaign on
PG\,1605+072.  Steward Observatory's contribution to MSST was 1.8\AA\
resolution spectra obtained during six clear contiguous nights on the
2.3~m Bok telescope on Kitt Peak.  Due to a combination of observing
technique and data reduction methods, the noise in the RV power
spectrum for PG\,1605+072 from the single-site Bok data was well under
1~km~s$^{-1}$, lower by almost a factor of two (Green et al., in
preparation), than was achieved by O'Toole et al.\ (2002).  With a
much longer time baseline for PG\,1627+017, we expected to be able to
detect any g-modes having amplitudes greater than
$\sim$1~km~s$^{-1}$.

As noted above, PG\,1627+017 has the same brightness as PG\,1605+072.
For the latter, exposures with the 2.3~m spectrograph were held to
45~s in order to adequately sample the main 365--530~s periods.  With
the same telescope and instrument setup, 200~s integrations for
PG\,1627+017 would allow a factor of two improvement in both the
signal-to-noise (S/N) and the velocity errors, yet the pulsational
periods would be sampled 3--4 times better.  All else being equal, the
increased exposures and improved sampling should largely compensate
for the factor of six reduction in the observed photometric amplitudes
of PG\,1627+017 relative to PG\,1605+072, given a suitably longer time
baseline.

Unfortunately, this is not the whole story.  All previous
spectroscopic mode detections in sdB pulsators were found in p-mode
pulsators, and it was not clear what to expect for g-modes.  The
problem with linear pulsation theory is that, precisely, it is {\bf
linear} and thus no information is provided about amplitudes of
modes. Only nonlinear theory is able, in principle, to provide
information on the expected amplitudes, but, in practice, the
usefulness of nonlinear theory is quite limited.

While the motions in p-mode pulsators are predominantly in the radial
direction, they are in the horizontal direction for g-mode
pulsators. Linear theory does not provide estimates of amplitudes per
se, but it gives a formula to estimate the ratio of the horizontal to
radial velocity amplitudes (see, e.g., equation 7.3 of Unno et
al. 1979).  As applied to the dominant $\sim$480~s mode in
PG\,1605+072, this formula indicates that the ratio of radial to
horizontal velocities is about 17:1. If we now consider a long period
g-mode in PG\,1627+017, say 7000 s, this ratio goes down to
1:15. However, this does not mean that we should scale the expected
velocity excursions by these amounts, and conclude that velocity
variations will not be seen in PG\,1627+017. Indeed, one can hope to
pick up the ``horizontal'' motions near the limb, which would
translate into radial velocity variations.  This is precisely what has
been done by van Kerkwijk, Clemens, \& Wu (2000) and Thompson et al.\
(2003), who obtained time-resolved spectroscopy of G29--38, a
pulsating (g-mode) ZZ~Ceti star.  Against perhaps naive expectations,
they were able to measure velocity variations of several km~s$^{-1}$
associated with g-mode motions.  Furthermore, they succeeded with a
white dwarf, where the expected ratio of radial-to-horizontal
velocities is more like 1:250.

\section{Observations}

We were allocated 40 nights for time-series spectroscopy of
PG\,1627+017: 24 nights on the SO 2.3~m Bok telescope (March--July), a
week on the SAAO 1.9~m telescope in May, and 9 nights on the SSO 2.3~m
telescope, also in May.  The SAAO and the SSO sites were badly
affected by winter clouds on a majority of their nights.  Much more
unexpectedly, half of the SO nights were also clouded out, and several
other nights were affected by significant cirrus.  In addition, the
June SO nights suffered a noticeable drop in atmospheric transparency
due to smoke from nearby wildfires.  Tables~1 and 2 list the details
for all scheduled nights on each separate telescope.  104 hours of
observations from 21 usable telescope nights resulted in 1473 spectra
between mid-March and early June.

A paramount consideration for a multisite campaign is the accuracy of
the UT time in the image headers from different observatories.  The
time systems at all sites were considered to be accurate to within one
second, thus the main uncertainty was whether or not there was any
significant delay between the UT time stamp and the actual moment the
shutter opened.

\subsection{Steward Observatory (SO)}

The Boller \& Chivens Cassegrain spectrograph at Steward's 2.3~m Bok
telescope was used with an 832/mm grating in second order and a 1~mm
Schott 8612 blocking filter.  In this instrumental configuration, the
dispersion is 0.7\AA /pixel over a wavelength range of 3755--4595\AA,
giving a spectral resolution of 1.8\AA\ (R = 2315) with a
1.5$^{\prime\prime}$ slit.  The 1200x800 pixel Loral/Fairchild/Imaging
Systems CCD was thinned, packaged, and backside treated by the Steward
Observatory CCD Laboratory for enhanced blue response (80\% at
3700\AA, 95\% at 4000\AA).  The CCD gain and readnoise are
$\sim$2.1~e$^-$/ADU and $\sim$5.5~e$^-$, respectively.  We used
on-chip binning of 3~pixels in the spatial direction, and further
windowed the CCD to 1200x135 pixels to reduce the overhead time
between spectra to 12~s.  The time stamp comes from a GPS server at
the telescope, and is accurate to within a few milliseconds.  We
conducted extensive engineering tests to determine the time lag
correction between the ccd shutter and the UT time stamp in the image
header and verify its repeatability (to within 0.2~s).  We also
determined that there was no detectable flexure in the wavelength
direction for hour angles up to $\pm$5~hours at constant declination.

Spectra of PG\,1627+017 were usually taken in approximately hour-long
sets of sixteen 200~s exposures, with 60~s HeAr calibration arcs
before and after.  Prior to each set, we rotated our slit so that it
would be at the parallactic angle at the midpoint of the set, to
minimize light loss due to atmospheric dispersion.  At the same time,
we also adjusted the collimator focus, which varies with dome
temperature, to maintain the same FWHM of the arc lines throughout the
night.  During program exposures, we hand-guided the star on the slit,
since we felt that the autoguider in use at the time was unable to
maintain sufficient centering for uniform illumination of the slit,
possibly leading to systematic velocity effects.

During the April and June runs (especially the latter, because of
smoke from wildfires), we often used shorter sets for PG\,1627+017,
alternating with exposures of a nearby constant luminosity sdB,
PG\,1432+004, in order to measure the atmospheric extinction.
PG\,1432+004 was observed over the same or a slightly larger range of
airmass than our target star.

\subsection{South African Astronomical Observatory (SAAO)}

We used the 74$^{\prime\prime}$ (1.9~m) Cassegrain spectrograph with a
1200/mm grating in first order.  Spectra were projected onto a SITe
CCD that was prebinned by 2~pixels in the spatial direction, and
windowed to 1792x256 pixels.  The gain and readnoise are 0.9~e$^-$/ADU
and 5.94~e$^-$, respectively.  The time lag between the UT time stamp
and the shutter opening is less than a few milliseconds for this
system.  With a dispersion of 0.5\AA /pixel and a 1.5$^{\prime\prime}$
slit, we obtained a spectral resolution of 1.0\AA\ over the selected
wavelength range 3680--4490\AA.  Spectra were taken sequentially in
frame transfer mode, with 600~s integrations per image bounded by 60~s
CuAr arcs.

The SAAO observations were the only ones not observed at the
parallactic angle, since the spectrograph slit is aligned east-west
and not easily rotated.  Without an atmospheric dispersion
compensator, it was considerably more difficult to center the blue
image on the slit using only a red acquisition TV and guider.

\subsection{Siding Spring Observatory (SSO)}

The Double beam spectrograph (DBS) with a 1200/mm grating in second
order was used with a 1752x532 SITe CCD, binned 1x4 and windowed to
1752x113 pixels.  Only the blue spectra were used in the following
analysis. The gain and readnoise are 1~e$^-$/ADU and 6~e$^-$,
respectively.  The time stamp accurately reflects the shutter opening
time. The 0.5\AA /pixel dispersion and 2$^{\prime\prime}$ slit gave a
spectral resolution of 1.0\AA\ over the useful wavelength range
of 3680--4444\AA.  The SSO 2.3~m telescope has an alt-az mount, and
therefore the slit was always aligned at the parallactic angle.

The DBS and the CCD imager were mounted on opposite sides at the f/18
Nasmyth focus, allowing a quick switch between spectroscopy and
photometry according to weather conditions.  Spectra were taken when
conditions were too poor to obtain photometry relative to
reference stars in the same image, although that turned out to include
the majority of the available observing time for PG\,1627+017.  The
spectroscopic exposure times were adjusted to obtain a satisfactory
S/N ratio, averaging about 200~s, and the images were bounded by 60~s
TiAr arcs.

\section{Reductions}

The spectra were bias subtracted, flat-fielded, and cleaned of cosmic
rays in the standard manner.  All subsequent data reduction was done
with IRAF\footnote{The Image Reduction and Analysis Facility, a
general purpose software package for astronomical data, is written and
supported by the IRAF programming group of the National Optical
Astronomy Observatories (NOAO) in Tucson, AZ.} tasks at Steward,
including background removal and extraction to the one dimensional
spectra, in order to have complete homogeneity between all of the
different data sets.  The calibration arc identification and
wavelength dispersion corrections were done with the IRAF {\bf
onedspec} package.  We used an appropriate combination of each
observatory's line lists, the IRAF package datafiles idhenear.dat and
cuar.dat, and the MIT Wavelength Tables (Phelps 1982), to find an
accurate wavelength for every identifiable unblended line in the
spectrum that was sufficiently above the noise threshold.  The one
dimensional spectra from each observatory were calibrated using
exactly the same set of arc lines for every spectrum.  The dispersion
solutions were determined by linear interpolation between the
preceding and following arc exposures according to the UT times at
mid-exposure.  The wavelength-corrected spectra were then resampled
onto a logarithmic wavelength scale, in preparation for the RV
cross-correlations.  The continuum was removed by fitting a cubic
spline, dividing by the fit, and subtracting 1.0 to get a mean
continuum level of zero.

We calculated the S/N per pixel for each spectrum, including CCD
readnoise as well as photon statistics errors from the star and sky.
A number of spectra from each observatory turned out to have very low
S/N, due to clouds.  For the following analysis, we discarded all
those with S/N $<$ 10.  This left 1161 SO spectra with a median S/N =
39, 184 SAAO spectra with a median S/N = 24, and 128 SSO spectra with
a median S/N = 18.

The PG\,1627+017 spectrum in Figure~2 shows the characteristic sdB
Balmer lines.  The 4471\AA\ and 4026\AA\ He I lines are much less
obvious, and most of the metal lines appear lost in the noise.
Nevertheless, we note that both helium and metal lines add
considerable power to the cross-correlations, even for the lower
resolution SO data.  Weak lines are more important still for the SAAO
and SSO data, where the higher resolution must compensate for lower
S/N.

\section{Radial Velocities and Orbital System Parameters}

\subsection{Measurements and Errors} 

Radial velocities were derived using the double precision version of
the IRAF task {\bf fxcor}\footnote{Available at
http://iraf.noao.edu/scripts/extern/rvx.pl}, which cross-correlates
object spectra against a template spectrum.  Our experience has been
that higher precision RV's are obtained from {\bf fxcor} if the
spectra and templates are all identically resampled onto a log scale
and continuum subtracted prior to the cross-correlation, rather than
within the {\bf fxcor} task itself.  Separate templates for each night
were constructed by combining a large number of the higher S/N spectra
for that night.  The higher the S/N of the template, and the more
closely its spectrum matches the object spectra, the better the
cross-correlation velocities will be.  For the most precise relative
velocities, a high S/N template of the star itself, taken with the
same instrumental setup, is preferable to a theoretical template or to
a high S/N template of a different star, because the pattern of weak
helium and metal lines varies so widely from one sdB star to another.

The Fourier filter parameters were chosen such that only narrow lines
and the sharp cores of the Balmer lines would contribute to the
cross-correlation.  Although the wide profiles of the hydrogen lines
visually dominate the spectra, the extra power gained by including
their wings in the cross-correlation does not seem to translate into
increased accuracy, presumably because the cross-correlation peak is
so much wider.  Our preferred filter parameters for the SO data were
derived from a number of trial-and-error tests using spectra of
constant velocity sdB stars (obtained during previous observing runs
with the same instrumental configuration).  Best results were obtained
using a ramp filter with values for IRAF's fxcor wavenumber
parameters, {\it cuton} and {\it fullon}, corresponding to
approximately the Balmer line FWHM and half of that FWHM,
respectively.  The optimum values for the fxcor {\it cutoff} and {\it
fulloff} parameters were nearly equal, corresponding to a pixel value
somewhat smaller than the instrumental resolution, as measured by the
FWHM of the arc lines.  The optimal parameters for the SAAO and SSO
spectra were determined in a similar manner.  We fit the
cross-correlation peaks with a gaussian function, since this results
in higher precision than the other available options (parabola,
center1d, etc).  (We note that while the single precision fxcor
gaussian fitting routine sometimes fails to find a good solution, the
double precision version always works well.)  The Ca~II K line region
near 3933\AA\ was excluded from the cross-correlation, to avoid
potential interstellar contamination and/or imperfect background sky
subtraction (most of the spectra were taken near full moon).

The final velocities were derived with an iterative procedure.  We
constructed rough templates by median filtering the higher S/N spectra
for each night, weighted by the number of counts in each spectrum
prior to continuum subtraction.  The individual spectra were
cross-correlated against this initial template.  An improved template
was then constructed, after Doppler-shifting to remove the derived
velocities.  Pixellation was not an issue, since the star's orbital
velocity shifted the lines over multiple pixels.  A few iterations
were sufficient for convergence.  Finally, we determined the
night-to-night zero point differences, separately for each
observatory, by cross-correlating the final velocity templates from
each night against each other.

The IRAF fxcor routine outputs velocities and velocity errors.  It is
important to note that fxcor's errors are correct only to within a
scaling factor, which depends on the number of counts in the spectra
and the Fourier filter parameters used.  Armandroff, Olszewksi, \&
Pryor (1995) outlined a procedure using Monte Carlo simulations
covering the entire range of S/N in the data to determine the
appropriate scaling factors needed to derive the formal errors.  While
this procedure is invaluable for investigations involving many
different stars, it is extremely time consuming and, by its nature,
only addresses the internal errors, not any systematic errors.  We
chose instead to use the observed scatter of the PG\,1627+017
velocities to estimate the actual errors, as described in the
following sections.

\subsection{Orbital Parameters} 

In order to search for pulsational velocity variations in a star that
is a member of a binary, the orbital velocity must first be
subtracted. We began with the assumption of a circular orbit, as usual
with post-common envelope binaries.  We used five sets of velocities
observed over a 7.5 year time interval to derive the orbital solution:
the three sets of SO, SAAO, and SSO (2003) velocities discussed above,
Morales-Rueda et al.'s (2003) velocities from data taken in
2000--2001, and the velocities listed in Table~3, derived from 1\AA\
resolution MMT spectra going back to 1996.  The MMT velocities are
part of another ongoing program to study the binary properties of a
representative sample of sdB stars (Green et al.\ 2004); the zero
point of the MMT velocities is known to be within 2--3~km~s$^{-1}$ of
the standard IAU radial velocity system.  Since the SO, SAAO, and SSO
cross-correlations provided only relative velocities, a separate zero
point term for each set of velocity data was included in the solution,
to ensure that all the data were on the velocity system defined by the
MMT data.

The period search algorithm that we employed uses the Singular Value
Decomposition (SVD) method (Press et al.\ 1992) to minimize the
goodness-of-fit parameter $\chi^{2}$ for various sine curves over a
range of periods.  The orbital solution requires an appropriate weight
for each velocity, which is normally the inverse square of its error.
Without knowing the SO, SAAO, and SSO error scale factors at the
start, we used an iterative procedure to determine the orbital
parameters.  We derived an initial solution by fitting a sine curve to
the Morales-Rueda et al.\ and MMT velocities only, for which the
errors were already known.  We then subtracted the orbital velocity
contribution from the SO, SAAO, and SSO velocities, and used the
standard deviation of the velocity residuals as a first estimate of
the mean error for that site.  The ratio of the estimated error to the
average fxcor error provided the necessary error scaling factor.  The
advantage of this procedure is that it takes into account both
internal and systematic errors.  We rederived our orbital solution
using all five sets of weighted velocities, and then recalculated the
standard deviations of the SO, SAAO, and SSO velocity residuals to
find an improved scale factor.  The resulting periodogram gives a best
fit orbital period of 0.8292056 $\pm$ 0.0000014 days.  The derived
sinusoidal orbital velocity curve is plotted in Figure~3 together with
all of the observed velocities, folded with the orbital period.  The
velocity residuals for each individual night, after subtracting the
orbital motion, are shown in Figure~4.

Inspection of figures 3 and 4 reveals some problems that merit further
discussion.  First, the SAAO data (green dots) are
disappointingly noisy, and second, systematic trends with time in the
SSO data (red dots) apparently conspired to suggest a smaller orbital
velocity amplitude than is indicated by any of the other velocity
data.  Fortunately for the orbital parameters, it turns out that
neither the SAAO nor the SSO data set had a measurable effect on the
solution.  This is partly due to the error scaling procedure described
above, which resulted in very small weights for those spectra.  The
typical (median) errors derived using the velocity scatter about the
theoretical curve were 6.6~km~s$^{-1}$, 16~km~s$^{-1}$, and
24~km~s$^{-1}$ for the SO, SAAO, and SSO velocities, respectively.
In any case, there were so few spectra from either of those two sites
relative to the large number of contemporary SO spectra, that the
orbital solution would have been negligibly affected with any
acceptable choice of error scaling parameter.  We will return to the
question of the appropriate errors and the weighting in the following
section.

The derived parameters for PG\,1627+017 are listed in Table~4.  The
ephemeris for time T$_0$, defined to be the time when the sdB star is
at inferior conjunction, is

\smallskip
HJD(T$_0$) = (2\,452\,804.7723 $\pm$ 0.0034) + (0.8292056 $\pm$ 0.0000014) $\times$ E
\smallskip

\subsection{Nature of the Unseen Companion} 

The times of superior conjunction (orbital phase = 0.5), combined with
the photometric light curves, are sufficient to show that PG\,1627+017
is not an eclipsing system.  The stellar radius and rotational
velocity given in Table~4 assume a typical sdB mass of 0.49$M_\odot$
and tidally locked components.  The 2MASS JHK colors, J$-$H = $-$0.107
and J$-$K = $-$0.068, do not permit a main sequence secondary with a
mass greater than 0.4--0.5$M_\odot$ (Green et al.\ 2005).  The orbital
period and derived mass of the unseen companion, M$_2$ $>$~0.257
$\pm$~0.011\,$M_{\sun}$, suggest that the secondary is a white dwarf,
especially since there are no known sdB + dM binaries with periods
longer than 0.33~days and minimum secondary masses greater than
0.2$M_{\sun}$.  Additionally, we know that a main sequence companion
with M$_2$ $>$ 0.25$M_{\sun}$ in an orbit shorter than one day should
produce a small but observable reflection effect (Maxted et al.\
2002).  The pulsational brightness variations in PG\,1627+017 would
almost certainly overwhelm a small reflection effect for any visual
inspection, but the reflection should be clearly detectable as a 0.83
day (13.96~$\mu$Hz) periodicity in a photometric power spectrum.
Randall et al.\ (2005b) did not note such a periodicity in the results
of their seven week photometry campaign, primarily because their
flattening procedure to correct for differential atmospheric
extinction removed nearly all power at frequencies longer than 0.1
$\mu$Hz.  We can now report that a reexamination of the power spectrum
using Randall et al.'s {\it unflattened} PG\,1627+017 light curves
reveals no evidence for any peak near 13.96~$\mu$Hz, even probing well
below the minimum 3$\sigma$ noise threshold.  Since there is no
reflection effect, we conclude that the secondary companion in
PG\,1627+017 is indeed a white dwarf.


\section{Time-Series Analysis}

Since the quality of our velocities varies considerably between
different sites, from night to night, and even within nights at a
single site, it was necessary to use a weighted Fourier Transform
technique, such as described in Kjeldsen \& Frandsen (1992).  This
requires revisiting the error scaling.  Weighting by the previously
derived errors effectively reduces the contribution of the SSO and
SAAO velocities to almost zero, which negates any advantage that might
be obtained from more complete longitude coverage.  Although the
scaled errors for the SAAO data are not too unreasonable, in view of
the velocity scatter in both Figures~3 and 4, the very large
systematic errors in the SSO velocities (the red dots in Figure~3) are
not appropriate weighting factors for investigating velocity
variations on time scales shorter than one night.  The SSO scatter
about their nightly trends in Figure~4 is nearly as good as the
scatter in the SO data. Therefore, we used an alternate method of
error scaling for the input to the Fourier transform, in which we
detrended the data in Figure~4 by subtracting a linear fit from each
night's light curve, prior to determining the standard deviations. In
other words, the mean errors now correspond to the scatter about the
nightly trend.  With this scheme, the median scaled errors for the
SO, SAAO, and SSO data sets are 5.2~km~s$^{-1}$, 9.5~km~s$^{-1}$, and
6.2~km~s$^{-1}$.

As mentioned earlier, complete longitude coverage is particularly
valuable for long-period sdB variables, to reduce aliasing.  However,
using all the data listed in Table~1 resulted in a rather noisy power
spectrum, so we experimented with a number of different subsets of the
data.  After trying numerous combinations, we concluded that the best
longitude coverage with the lowest noise was obtained from the subset
comprised of the last two SSO nights, the last two SAAO nights plus
two short sections of the second (HJD between 774.4 and 774.52) and
third SAAO nights (775.48 to 775.56), and all of the SO data except
the first March night and the first quarter of the fifth SO night.

Figure~5 shows the amplitude spectrum for the best data set, which is
comprised of 1285 spectra representing 84 hours of spectroscopy over a
53 day time period.  (Note that any uncertainties due to the already
low duty cycle of 6.6\% must be exacerbated by the nonuniform
distribution of the observations in both longitude and time.) The top
panel shows several weaker peaks near the middle, representing excess
power in the frequency range between 120 and 160~$\mu$Hz, as well as a
few stronger peaks at very low frequencies (discussed below).  By
design, we obtained our spectroscopic observations during the same
April--June 2003 time period covered by Randall et al.'s (2004, 2005b)
photometric campaign on PG\,1627+017.  The midrange peaks in Figure~5
fall in the same 120 to 170~$\mu$Hz range as the strongest of the
photometric pulsation modes, the first indication that our data have
reached the level where g-modes can be spectroscopically detected
in this star.  There is no further structure to be found at
frequencies above 375~$\mu$Hz as shown in the top panel of Figure~5.

The second panel in Figure 5 is a slightly expanded view of the top
panel, without the low frequencies.  The window function for one of
the peaks is plotted in the third panel, with insets showing the
details of the excess power structure.  The bottom panel shows the
result of pre-whitening ({\it e.g.}\ Bill\`eres et al.\ 2000) by the
strongest peak (f1) at 138.87~$\mu$Hz.  In the pre-whitening
procedure, the full Fourier Transform of the lightcurve was
calculated, and a peak-finding algorithm was used to determine the
frequencies of the peaks rising above a specified threshold in the
Fourier domain, assumed to be three times the mean level of the
background noise.  The frequency of the highest amplitude peak was
used as a fixed parameter in a nonlinear least-squares fitting
procedure to find the corresponding amplitude and phase that best fit
the lightcurve.  The sine-wave generated for that frequency,
amplitude, and phase was subtracted from the lightcurve, removing the
strongest peak as well as any associated sidelobes.  The FT was then
recalculated.  If any other peaks had still remained above the chosen
noise threshold, the procedure would have been repeated to remove the
next strongest frequency, and so on, until no more significant peaks
were left.  In our case, a single iteration suffices, because only one
peak appears to be significant.  We will return to this point
following a brief discussion of the low frequencies.

When calculating the Fourier transform (FT), it is important to start
with a lower frequency limit of zero.  Otherwise, if there is real
power at very low frequencies that isn't properly subtracted, the
corresponding sidelobes at higher frequencies might affect the
subsequent analysis.  Typically, the lowest frequency peaks are simply
assumed to result from longterm instrumental drifts, and often they
can be simply ignored.  Instrumental drifts do occur in our data, as
seen in Figure~4, but they are not responsible for the strongest low
frequency effect in the top panel of Figure~5.  The central peak of
the most obvious low frequency window function pattern turns out to be
27.910~$\mu$Hz, exactly half of the orbital period (27.916~$\mu$Hz),
to within the errors.  The reality of this peak, and its frequency, are
indisputable, as they were derived and corroborated in three
independent analyses.

Edelmann et al.\ (2005) find similar radial velocity distortions in at
least three, and possibly five, other sdB stars, which they conclude
must be evidence for slightly elliptical orbits.  Short-period sdB
binary systems like these are considered to be the result of mass
transfer followed by common envelope ejection.  Even small residual
eccentricities would be quite surprising, as it is widely expected
that the components will emerge from the common envelope phase with
completely circularized orbits (e.g.\ Terman, Taam \& Hernquist 1994;
also Taam, personal communication).  A theoretically more palatable
possibility would be another light source within the binary (e.g.\ the
unseen companion or its heated face) whose changing Balmer absorption
line spectrum, as seen by an earth observer, might distort the radial
velocity curve from being purely sinusoidal.  However, as noted above,
no contribution at this frequency is detected in Randall et al.'s
simultaneous photometry, down to a small fraction of a percent, and
the photometry is considerably more sensitive than our spectroscopy.
Furthermore, we were unable to reconcile the phases of the half-orbital
term and the orbital ephemeris with what would be expected from any
type of light variation mechanism around the orbit.  More work is
clearly needed to clarify this problem.

Regardless of the underlying mechanism, including in our orbital
solution a small harmonic term with twice the orbital frequency
removed the majority of the low frequency structure (Figure~6),
without detectably altering the frequency or amplitude of the main
higher frequency peaks.  The remaining very low frequency peaks can be
reasonably attributed to the observed instrumental drifts, as seen in
Figure~4, and their removal does not affect the 120--160~$\mu$Hz
frequencies either.

The highest peak in the pulsational range in Figure~6 (top two panels)
is the one at 7201.0~s (138.87~$\mu$Hz).  Its amplitude,
1.44~km~s$^{-1}$, is almost 4$\sigma$ above the mean noise level of
0.365~km~s$^{-1}$.  As stated earlier, prewhitening by this frequency
leaves no other significant peaks above the noise level.  However, at
the same time that our peak-finding algorithm located the 138.87
$\mu$Hz peak, it also found two other nearby peaks with only slightly
weaker amplitudes, about 1.23~km~s$^{-1}$, at frequencies of 7014.6~s
(142.56~$\mu$Hz) and 7037.3~s (142.10~$\mu$Hz).  The amplitudes are
sufficiently comparable that we cannot be sure which of the three
peaks is the truly significant one.  It could well be the one with the
formally highest amplitude, or it might be one of the others.
Separate trials showed that pre-whitening by either of the other two
frequencies was also sufficient to reduce all the other peaks below
the 3$\sigma$ noise level.  To assess the possibility that peaks of
these amplitudes might occur by chance, we calculated the false alarm
probability (Scargle 1982; Horne $\&$ Baliunas 1986).  Using the
observed noise level of 0.365~km~s$^{-1}$, there is a 94\% chance that
the strongest peak is real, or about a 74\% chance for either of the
two weaker peaks.  Therefore, it seems highly probable that we have
detected at least one real periodicity, although the data are too
ambiguous to specify with any certainty which one it is.

The amplitudes of the three peaks vary somewhat if we use different
sets of input nights or different weighting, although the 7201~s peak
is always the strongest.  The three possible peaks are robust in the
sense that they always show up at the same periodicities, to within a
fraction of a second, for most combinations of input velocity data
that we tried, as long as the last two nights of SSO data are included
and assigned non-negligible weights.  Unfortunately, there is little
difference even when we discard the SAAO and SSO nights entirely,
which means that our campaign gained relatively little advantage from
the additional longitude coverage.  Using only the SO data with a low
frequency (``elliptical'') term included in the subtracted orbital
solution (Figure~7), the sidelobes in the window function are a little
stronger, marginally higher peaks are found at 7200.5~s and 7013.1~s,
although the 7037~s peak is no longer distinguishable, and the noise
is slightly increased to 0.374~km~s$^{-1}$.

It is instructive to compare the frequencies of the photometric and
spectroscopic detections.  Randall et al.'s (2004, 2005b) photometric
analysis uncovered a fairly rich pulsational amplitude spectrum with
the strongest peak (amplitude 0.49\%) at 6630~s (150.83~$\mu$Hz), and
the next two strongest peaks at 6664~s (150.06~$\mu$Hz) and 7035~s
(142.15~$\mu$Hz) (amplitudes 0.39\% and 0.38\%, respectively).  The
one day alias of our 7201~s peak is 6646~s (150.48~$\mu$Hz), which
happens to lie halfway between the two strongest and closest
photometric peaks.  Furthermore, the other two potential detections at
7014~s (142.56~$\mu$Hz) and 7037~s (142.10~$\mu$Hz) are comparably
close to the next strongest photometric peak in the photometry.  Could
the 7201~s spectroscopic peak be an undistinguished blend, and are the
spectroscopic uncertainties large enough to identify our peaks with
the photometric detections?  The frequencies derived from the Fourier
transform have no formal error estimates attached to them, but
Kjeldsen's (2003) eqn.\ 3, $\sigma _f$ = 0.44 * 1/SNR * 1/T, together
with our peak SNR of about 3.9, gives an error that is about 1/10 of
the time resolution.  The nominal spectroscopic time baseline of 53
days would therefore suggest a very tiny error of only $\sim$0.02
$\mu$Hz.  However, Kjeldsen's equation does not include the
uncertainty from beating between closely spaced modes in the star or
the effect of the window function, which is demonstrably very complex
(as seen in the third panel insets in Figures~5, 6, and 7) due to our
extremely low duty cycle.

In any case, the likelihood that all of the three possible
spectroscopic peaks should correspond so closely with the three
strongest photometric peaks purely by chance is very remote.  It seems
obvious that we have recovered one or more of the photometric modes,
considerably strengthening the argument for a spectroscopic detection
of g-mode pulsation in this star.

\section{Amplitude Variations as a Function of Wavelength}

Theory predicts that luminosity variations in sdB stars should be
wavelength dependent, with the largest amplitudes at the positions of
the Balmer lines and at bluer continuum wavelengths, especially toward
the ultraviolet (see, e.g., Randall et al.\ 2005a).  Although we
suspected that our modest wavelength range (3755--4595\AA) would be
too small to show much of a continuum effect, we nevertheless acquired
spectra for the nearby, apparently constant luminosity sdB star,
PG\,1432+004, in order to correct for atmospheric extinction as a
function of wavelength.  PG\,1432+004 (14$^h$35$^m$19$^s$.8,
+00$^{\degr}$13$^{\arcmin}$52$^{\arcsec}$) was observed only at the
Steward Observatory 2.3~m telescope, in April and again in June when
the extinction was problematic due to smoke from fires.  PG\,1432+004
rises and sets two hours ahead of PG\,1627+017, and has very nearly
the same brightness (V = 12.75) and declination, which made it very
easy to observe between sets of PG\,1627+017 observations.

To calculate the extinction, we divided the PG\,1432+004 spectra into 63
wavelength bins.  We normalized each spectrum by the counts at
4485--4585\AA, summed the counts in each wavelength bin, and fit a
straight line as a function of wavelength.  The extinction
corrections were applied to the normalized PG\,1627+017 spectra, 
which were then shifted to the rest wavelength before determining
the counts in their wavelength bins.

We computed power spectra for the bluest continuum region that could
be well measured ($\sim$~3950\AA), but as expected, the wavelength
difference between that point and the reference wavelength at the red
end of the spectrum was far too small to distinguish any pulsational
effects.

On the other hand, the higher order Balmer lines did exhibit
noticeable amplitude variations, that were weak or absent in H$\delta$
and H$\gamma$ (the longest wavelength Balmer lines in the spectra).
Figure~8 shows the FT's for three representative Balmer lines, H11
(top), H8 (center), and H$\gamma$ (bottom).  Unfortunately, only one
frequency corresponding to 7209~s (138.7~$\mu$Hz) could be convincingly
detected above the noise in any of the amplitude power spectra.  (Two
lower peaks, most easily seen in the central panel, slightly more than
10~$\mu$Hz to either side of the 7209~s peak, are sidelobes.)

While the observed amplitudes are not known to sufficient accuracy to
allow the determination of the $l$ index for the 7209 s mode, it is
still interesting to investigate if, at least, some constraints could
not be derived.  Randall et al.\ (2005a) have recently adapted the
theory of multicolor photometry to pulsating sdB stars, including
detailed nonadiabatic effects in the atmospheric layers. Following
that paper, we computed the expected relative amplitudes of a 7209 s
g-mode for a model of PG 1627+017 using the atmospheric parameters
listed in Table 4. Our results are presented in Figure~9 which shows
the expected monochromatic amplitude of the 7209 s mode relative to
that of the core of the H11 line at 3770\AA~ for different assumed
values of the degree index $l$. The points with error bars are our
measured line core amplitudes with respect to the amplitude of the H11
line. Not unexpectedly, the uncertainties on the amplitude ratios are
so large that we cannot discriminate between values of $l$ = 1, 2, 4,
or 6. Interestingly, however, our data are good enough to rule out the
possibility that the 7209 s mode has a value of $l$ = 3 or $l$ = 5.

\section{Summary and Conclusions}

We were allocated 40 nights during March--July 2003 on 2~m class
telescopes in Arizona, South Africa, and Australia to attempt the
first spectroscopic detection of gravity modes in a long-period
(g-mode) sdB pulsator.  Our equatorial target, PG\,1627+017, is one of
the brightest PG\,1716 variables, with moderately large amplitudes for
its type.  Previous photometry suggested pulsational periods of the
order of 1--2 hours and amplitudes $\lta$1\%.  For ground-based
observations, minimum time baselines of many weeks are required to
adequately resolve the modes, as well as good longitude coverage to
eliminate aliasing.  Order-of-magnitude calculations suggested that we
would need a detection limit $\lta$1~km~s$^{-1}$ and a velocity
precision about 4--5~km~s$^{-1}$, to set useful upper limits on
possible g-mode amplitudes.  Ideally, we wanted to obtain time-series
spectroscopy on 2~m class telescopes for a week of overlapping nights
at all three observatories during the middle of the corresponding
photometric campaign, plus additional blocks of 4--5 nights at a time
at Steward Observatory, every few weeks for at least a month or so
before and after the main effort.  Although such spotty coverage can't
provide very good resolution, it took a huge effort merely to obtain
this much 2~m telescope time and carry out the observations.  It would
be extremely difficult to get a higher duty cycle without dedicated
telescopes.  Unfortunately, bad weather limited the number of our
useful nights to a bare minimum of 18: a total of 11 nearly
overlapping May nights from the three sites, plus one additional April
night and five June nights in Arizona.

The final velocity data set included 84 hours of time-series
spectroscopy over a formal time baseline of 53 days.  We combined our
new velocities for PG\,1627+017 with published velocities from
Morales-Ruelas et al.\ (2003), and with newly reported MMT velocities
(Table~3), to derive improved orbital parameters (Table~4).

Surprisingly, the radial velocity power spectrum shows a significant
peak at 27.910~$\mu$Hz, which, to within the errors, is exactly half
of the orbital period (27.916~$\mu$Hz), corroborating Edelmann et
al.'s (2005) discovery of similar small velocity distortions in
several other short-period sdB stars.  Edelmann et al.\ conclude that
slightly elliptical orbits are the only possible explanation.  We are
not completely convinced that this is the case, but we could not find
any alternative explanations.  The simultaneous photometry and
spectroscopy campaigns for PG\,1627+017 constrain the possibilities even
more tightly, because the mechanism responsible for the half-orbital 
velocity term produced no detectable brightness variations.

In spite of an overall efficiency of only 6.6\% and a somewhat higher
detection limit than we had hoped, we appear to have reached the level
where g-modes can be observed in a sdB star.  The power spectrum of
the residual velocities, after subtracting the orbital motion, shows
excess power in the same 120--170~$\mu$Hz frequency range as the
observed photometric pulsations.  Three peaks are found at 7201.0~s
(138.87~$\mu$Hz), 7014.6~s (142.56~$\mu$Hz) and 7037.3~s
(142.10~$\mu$Hz), with amplitudes that are 3--4$\sigma$ above the mean
noise level of 0.365~km~s$^{-1}$.  At least one of the peaks is likely
to be real based on noise statistics alone, but all three are good
candidates because each one is tantalizingly close to, or a one day
alias of, one of the three strongest photometric peaks found by
Randall et al.\ (2004, 2005b).  As this is most unlikely to happen by
chance, we conclude that we have indeed detected g-mode pulsations in
PG\,1627+017 at the 1.0--1.5~km~s$^{-1}$ level.  This indicates that
the strongest velocity mode in this cool, low gravity, g-mode
pulsator, relative to its $\sim$0.03 mag photometric variations, is
only about a factor of two weaker than the strongest velocity mode
(14~km~s$^{-1}$) observed in the correspondingly cool, low gravity
p-mode pulsator, PG\,1605+072, relative to its 0.2 mag photometric
variations.

We further attempted to detect pulsational variations in the Balmer
line amplitudes.  The higher order Balmer lines, in particular, showed
evidence for amplitude variations with a periodicity of 7209~s, once
again quite close to the dominant photometric frequencies.  The trend
of the pulsational amplitudes with wavelength is consistent with with
theoretical expectations, and we were able to rule out the $l$ = 3 and
$l$ = 5 possibilities for that mode.

We conclude that spectroscopic detection of gravity modes in sdB stars
is (just) within the realm of possibility for the strongest PG\,1716
pulsators.  Truly useful results for velocity modes would require much
higher efficiency spectral monitoring over at least a comparable time
baseline (6--8 weeks), with a radial velocity accuracy closer to
1--2~km~s$^{-1}$.  Achieving this level of efficiency and precision
with ground-based facilities is very unlikely at present, since it
would require long blocks of telescope time on multiple telescopes
larger than 2~m, with good longitude coverage.  Since the necessary RV
accuracy would also be a problem for satellite observations, it
appears that spectroscopic mode detection in the PG\,1716 stars, as
well as in many EC\,14026 stars, has a way to go before it will be
truly practical.

\acknowledgments 

We thank the SO, SSO, and SAAO TACs for their very generous time
allocations.  This project was partially funded by NSF grants
AST-9731655 and AST-0098699, and also by Australian Research
Council. L.L.K is supported by a University of Sydney Postdoctoral
Research Fellowship. G.F. is supported by the Natural Sciences and
Engineering Research Council of Canada and by the Canada Research
Chair Program.

\clearpage

\clearpage

\begin{deluxetable}{ccccccr}
\tablecolumns{7}
\tablewidth{0pc}
\tablecaption{Spectroscopic log for PG\,1627+017 observations\tablenotemark{a}}
\tablehead{
\colhead{Start UT date} & \colhead{Begin HJD} & \colhead{End HJD} 
& \colhead{Observatory} & \colhead{Telescope} & \colhead{N$_{\rm spec}$} 
& \colhead{N$_{\rm hours}$} \\
\colhead{(year 2003)} & \colhead{(2452000+)} & \colhead{(2452000+)} & 
\colhead{} & \colhead{} & \colhead{(S/N $>$ 10)} & \colhead{}}
\startdata

15 March\dotfill & 713.849 & 714.033 & SO\tablenotemark{b} & 2.3~m &  53  & 2.94 \\
16 March\dotfill & -- & -- & SO & 2.3~m & ~~0 & --~~ \\
10 April\dotfill & -- & -- & SO & 2.3~m & ~~0 & --~~ \\
13 April\dotfill & -- & -- & SO & 2.3~m & ~~0 & --~~ \\
14 April\dotfill & -- & -- & SO & 2.3~m & ~~0 & --~~ \\
15 April\dotfill & -- & -- & SO & 2.3~m & ~~0 & --~~ \\
22 April\dotfill & 751.755 & 751.960 & SO & 2.3~m &  76  & 4.22 \\
23 April\dotfill & -- & -- & SO & 2.3~m & ~~0 & --~~ \\
12 May\dotfill & -- & -- & SSO & 2.3~m & ~~0 & --~~ \\
13 May\dotfill & -- & -- & SO & 2.3~m & ~~0 & --~~ \\
13 May\dotfill & 773.048 & 773.241 & SSO\tablenotemark{b} & 2.3~m & 25 & 1.75 \\
13 May\dotfill & 773.389 & 773.674 & SAAO\tablenotemark{b} & 1.9~m & 33 & 5.50 \\
14 May\dotfill & -- & -- & SO & 2.3~m & ~~0 & --~~ \\
14 May\dotfill & -- & -- & SSO & 2.3~m & ~~0 & --~~ \\
14 May\dotfill & 774.348 & 774.670 & SAAO\tablenotemark{c} & 1.9~m & 32 & 5.33 \\   
15 May\dotfill & -- & -- & SO & 2.3~m & ~~0 & --~~ \\
15 May\dotfill & 775.113 & 775.288 & SSO & 2.3~m & 45 & 2.50 \\ 
15 May\dotfill & 775.376 & 775.651 & SAAO\tablenotemark{c} & 1.9~m & 34 & 5.67 \\
16 May\dotfill & 775.679 & 775.989 & SO & 2.3~m &  116~ & 6.44 \\
16 May\dotfill & -- & -- & SSO & 2.3~m & ~~0 & --~~ \\
16 May\dotfill & 776.460 & 776.590 & SAAO\tablenotemark{b} & 1.9~m & 17 & 2.83 \\
17 May\dotfill & 776.664 & 776.978 & SO & 2.3~m &  114~ & 6.33 \\
17 May\dotfill & 777.015 & 777.264 & SSO & 2.3~m & 58 & 3.22 \\
17 May\dotfill & -- & -- & SAAO & 1.9~m & ~~0 & --~~ \\
18 May\dotfill & 777.683 & 777.971 & SO\tablenotemark{c} & 2.3~m & 79 & 4.39 \\
18 May\dotfill & -- & -- & SSO & 2.3~m & ~~0 & --~~ \\
18 May\dotfill & 778.343 & 778.651 & SAAO & 1.9~m & 42 & 7.00 \\
19 May\dotfill & 778.667 & 778.984 & SO & 2.3~m & 119~ & 6.61 \\
19 May\dotfill & -- & -- & SSO & 2.3~m & ~~0 & --~~ \\
19 May\dotfill & 779.368 & 779.599 & SAAO & 1.9~m & 26 & 4.33 \\
20 May\dotfill & 779.657 & 779.952 & SO & 2.3~m & 112~ & 6.22 \\
21 May\dotfill & -- & -- & SSO & 2.3~m & ~~0 & --~~ \\
10 June\dotfill & 800.671 & 800.969 & SO & 2.3~m & 76 & 4.22 \\
11 June\dotfill & 801.664 & 801.972 & SO & 2.3~m & 114~ & 6.33 \\
12 June\dotfill & 802.661 & 802.971 & SO & 2.3~m & 82 & 4.56 \\
13 June\dotfill & 803.655 & 803.972 & SO & 2.3~m & 120~ & 6.67 \\
14 June\dotfill & 804.654 & 804.969 & SO & 2.3~m & 100~ & 5.56 \\
04 July\dotfill & -- & -- & SO & 2.3~m & ~~0 & --~~ \\
05 July\dotfill & -- & -- & SO & 2.3~m & ~~0 & --~~ \\
06 July\dotfill & -- & -- & SO & 2.3~m & ~~0 & --~~ 

\enddata
\tablenotetext{a}{Steward Observatory = SO; South African Astronomical Observatory = SAAO; 
   Siding Spring Observatory = SSO}
\tablenotetext{b}{Data not included in RV power spectrum}
\tablenotetext{c}{Partial night's data included in power spectrum (see text)}
\end{deluxetable}

\clearpage

\begin{deluxetable}{ccccc}
\tablecolumns{5}
\tablewidth{0pc}
\tablecaption{Spectroscopic log for PG\,1432+004 observations, Steward Observatory only}
\tablehead{
\colhead{Start UT date} & \colhead{Begin HJD} & \colhead{End HJD} 
& \colhead{No.\, of spectra} & \colhead{No.\, of hours} \\
\colhead{(year 2003)} & \colhead{(2452000+)} & \colhead{(2452000+)}
&\colhead{(S/N $>$ 10)} & \colhead{}}
\startdata

14 April\dotfill & 743.684 & 743.859 & 64 & 3.56 \\
10 June\dotfill & 800.713 & 800.894 & 16 & 0.89 \\
12 June\dotfill & 802.702 & 802.895 & 20 & 1.11 \\
14 June\dotfill & 804.694 & 804.872 & 10 & 0.56

\enddata
\end{deluxetable}

\clearpage

\begin{deluxetable}{rrc}
\tablecolumns{3}
\tablewidth{0pc}
\tablecaption{MMT radial velocities (1996--2002)} 
\tablehead{
\colhead{HJD} & \colhead{V} & \colhead{V$_{\rm err}$} \\
\colhead{(2450000+)} & \colhead{(km s$^{-1}$)} & \colhead{(km s$^{-1}$)}}
\startdata

 243.810746 &  $-$72.81  &   1.48 \\
 260.762533 &  $-$50.61  &   1.59 \\
 287.754529 &  $-$71.29  &   1.34 \\
 626.801965 &  $-$19.73  &   1.20 \\
 627.760116 &  $-$89.81  &   1.26 \\
 642.768932 & $-$117.98  &   1.41 \\
2531.603374 &  $-$87.75  &   1.35 \\
2532.622904 & $-$126.22  &   1.35

\enddata
\end{deluxetable}

\clearpage

\begin{deluxetable}{cc}
\tablecolumns{2}
\tablewidth{0pc}
\tablecaption{Derived system parameters for PG\,1627+017 } 
\tablehead{
\colhead{Parameter} & \colhead{Value}}
\startdata

T$_{\rm eff}$   &  23669 $\pm$ 190 \\
log g           &  5.315 $\pm$ 0.021 \\
R ($R_{\sun}$)  & 0.255 $\pm$ 0.013 \\
Period (days)   & 0.8292056 $\pm$ 0.0000014 \\
T$_{0}$ (days)  & 2452804.7723 $\pm$ 0.0034 \\
K (km s$^{-1}$) & 70.10 $\pm$ 0.13 \\
$\gamma$ (km s$^{-1}$)       & $-$54.16 $\pm$ 0.27 \\
V$_{\rm rot}$ (km s$^{-1}$)  & 15.567 $\pm$ 0.823 \\
M$_{2,{\rm min}}$ ($M_{\sun}$)& 0.257 $\pm$ 0.011

\enddata
\end{deluxetable}


\clearpage

\begin{figure}
\epsscale{.80}
\plotone{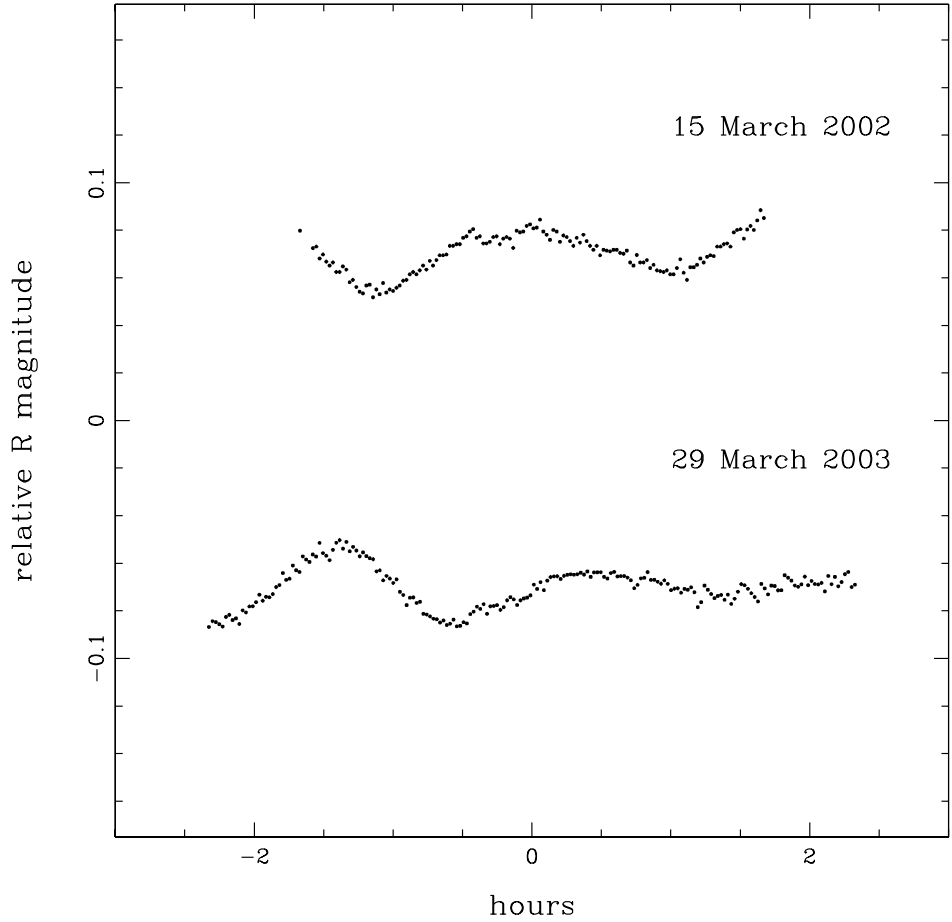}
\figcaption{Two typical light curves for PG\,1627+017 taken prior to the
beginning of our campaign, including the discovery light curve (top), 
illustrating amplitude variations that are relatively large for a
long-period sdB variable.
(The increased scatter at the end of the bottom light curve is due to 
thin cirrus.)\label{fig1}}
\end{figure}

\clearpage
\begin{figure}
\includegraphics[width=0.7\textwidth, angle=-90]{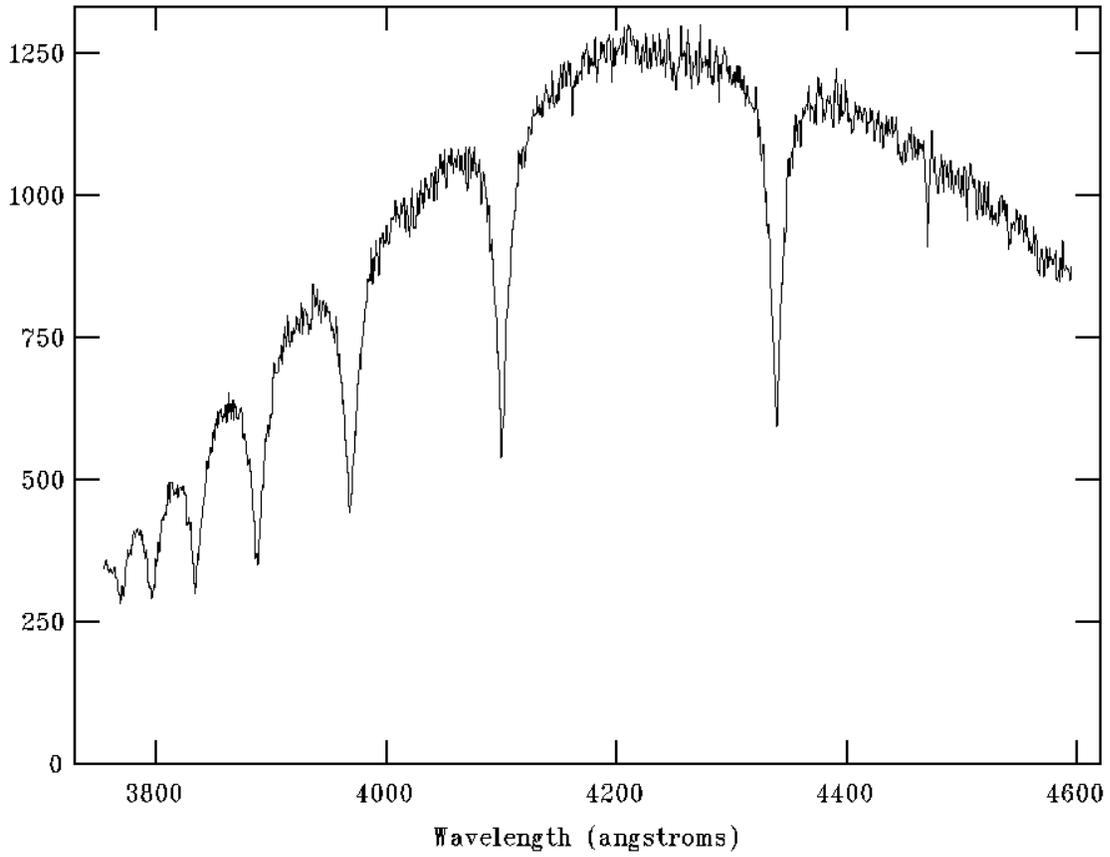}
\figcaption{A typical 1.8\AA\ resolution spectrum of PG\,1627+017 
taken at the Steward 2.3~m telescope.\label{fig2}}
\end{figure}

\clearpage
\begin{figure}
\epsscale{.80} 
\plotone{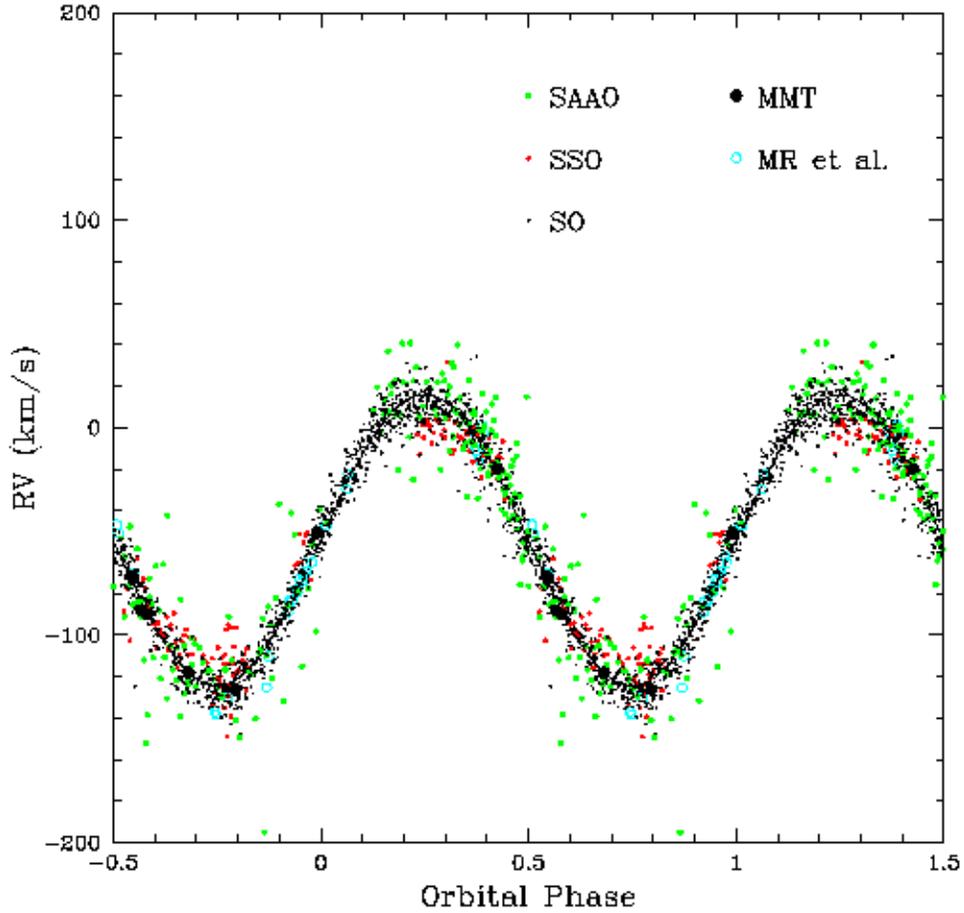} 
\figcaption{The radial velocity {\it vs} orbital phase plot for
PG\,1627+017, with the best fit theoretical RV curve for P =
0.8292056$^d$ superimposed. Included are the 1473 velocities derived
from SO, SAAO, and SSO spectra with S/N $>$ 10, plus additional
velocities from Morales-Rueda et al.\ (2003) and Table~3, for a total
time baseline of 7.5 years. \label{fig3}}
\end{figure}

\clearpage
\begin{figure}
\epsscale{.80}
\plotone{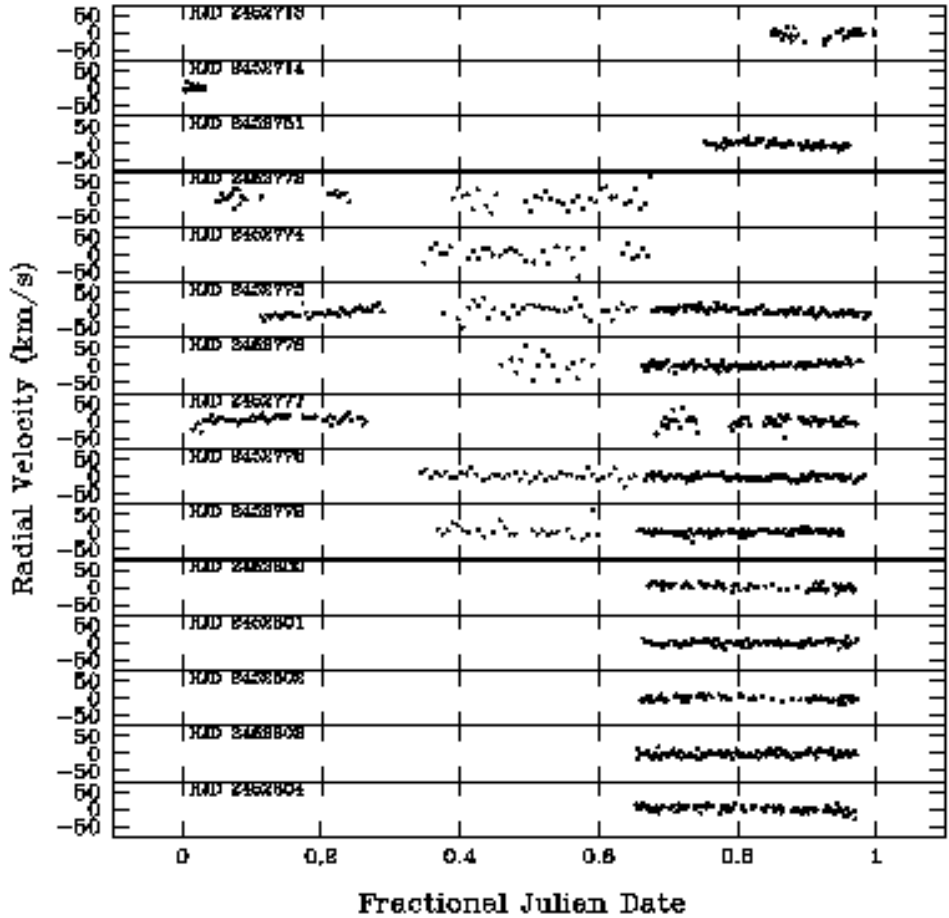}
\figcaption{Residual radial velocities for each night, after subtracting
out the orbital motion.  The SO data are on the right side of the plot
(except for the end of the first night, which wraps around to the top
left), the SAAO data are in the middle, and the SSO data are on the
left (see Table~1).\label{fig4}}
\end{figure}

\clearpage
\begin{figure}
\epsscale{.80}
\plotone{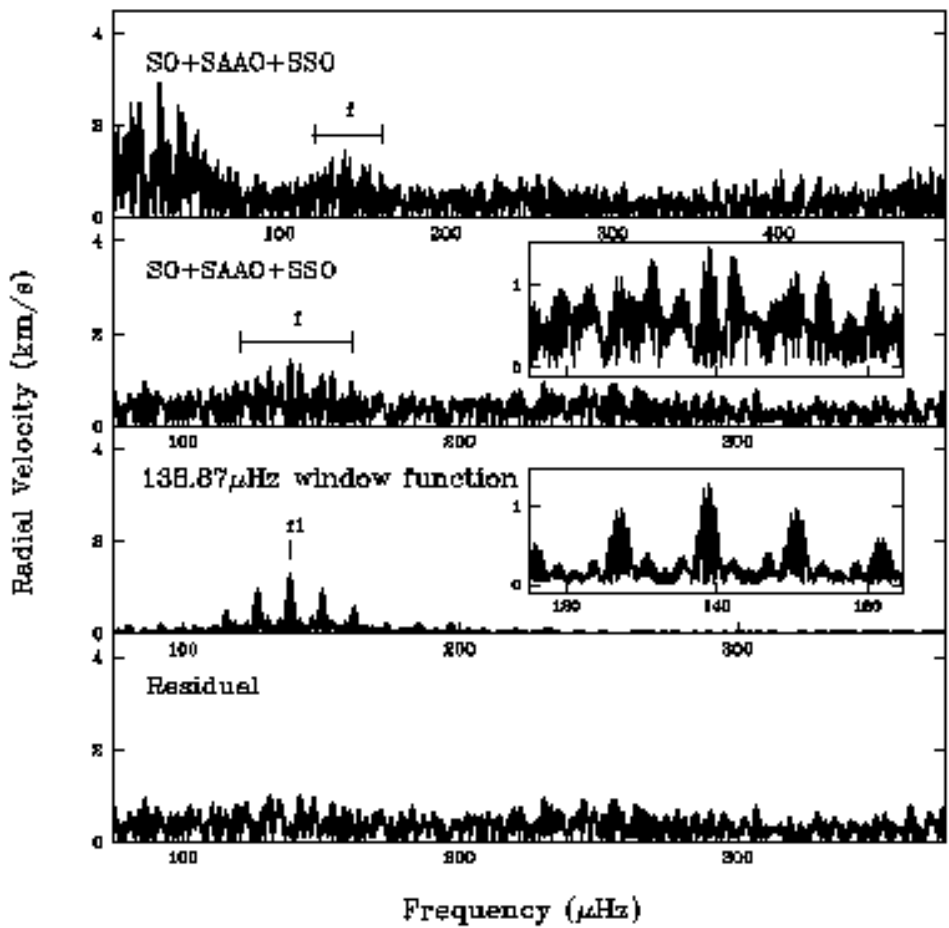}
\figcaption{The top panel shows the velocity amplitude spectrum in the
range from 0~$\mu$Hz--500~$\mu$Hz after removing the orbital frequency.
The second panel shows only the 75~$\mu$Hz--375~$\mu$Hz range, with the
subwindow giving an expanded view of the region of excess power (f)
between 120 and 160~$\mu$Hz.  The third panel is the window function
for a single pulsation frequency of 138.87~$\mu$Hz sampled at the
observed HJD, with a subwindow on the same scale as the one above.
The bottom panel is the residual amplitude spectrum after removing the f1
frequency.\label{fig5}}
\end{figure}

\clearpage
\begin{figure}
\epsscale{.80} 
\plotone{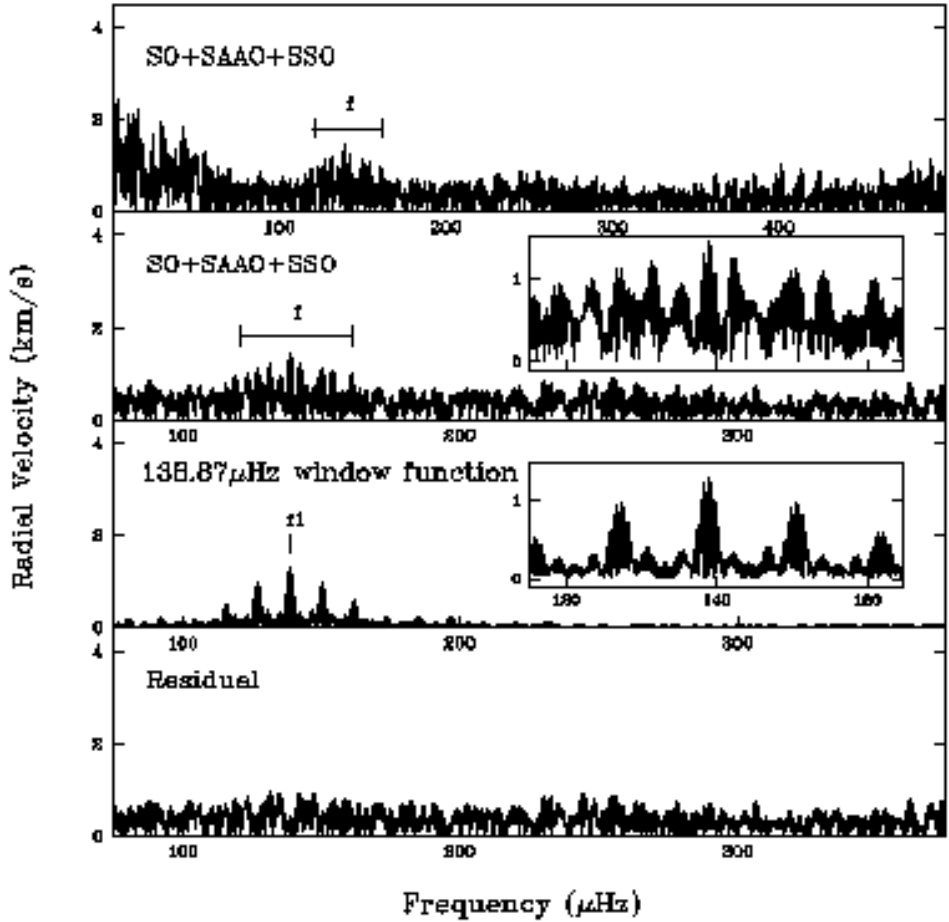} 
\figcaption{Same as Figure~5, except
that the input velocities were calculated using an orbital solution
that includes a small ellipticity term.  Note that the strongest low
frequency peaks (centered at half the orbital period, 27.92~$\mu$Hz in
Figure~5) have now disappeared, but the peaks corresponding to the
range of photometric pulsation frequencies are essentially
unchanged.\label{fig6}}
\end{figure}

\clearpage

\begin{figure}
\epsscale{.80}
\plotone{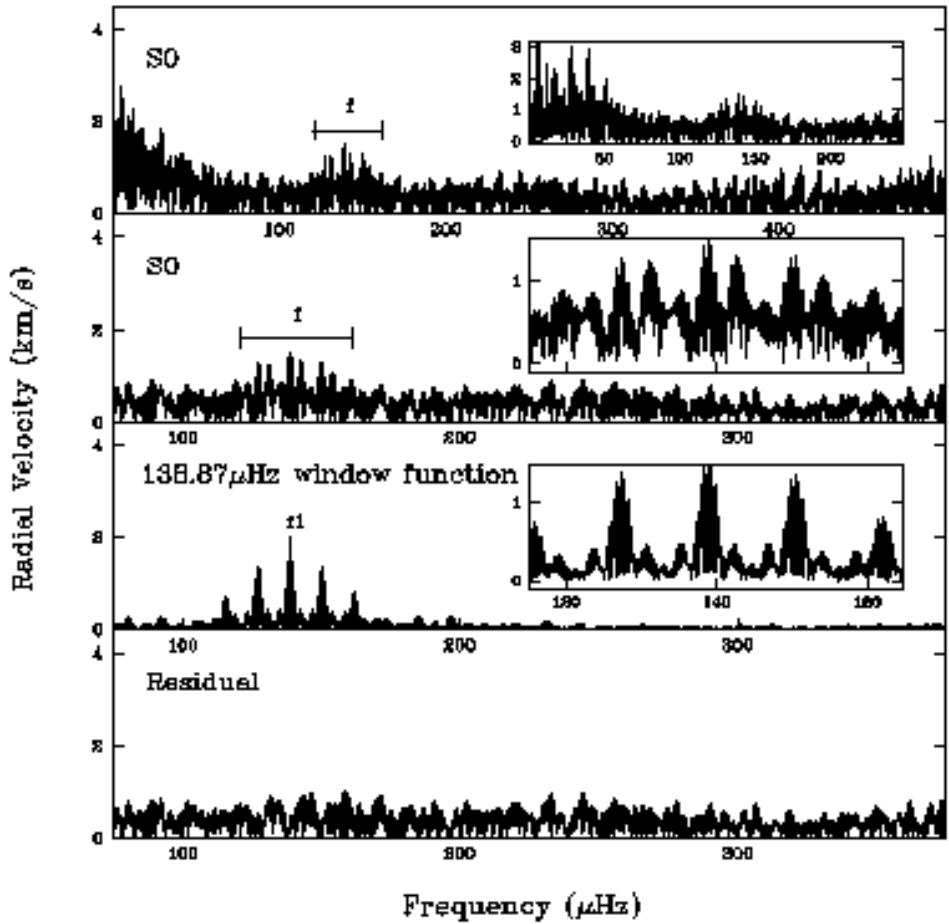}
\figcaption{Similar to Figure 6, except that only the April--June SO data
were used to construct the power spectrum.  The noise is slightly larger
here and the structure in the 120--160~$\mu$Hz region somewhat changed,
but the differences are relatively minor.  The subwindow in the top panel
shows an expanded view of the low frequency region prior to the removal of 
the half-orbital term, which shows up more cleanly in the SO data alone
than in Figure~5.\label{fig7}}
\end{figure}

\clearpage
\begin{figure}
\epsscale{.80}
\plotone{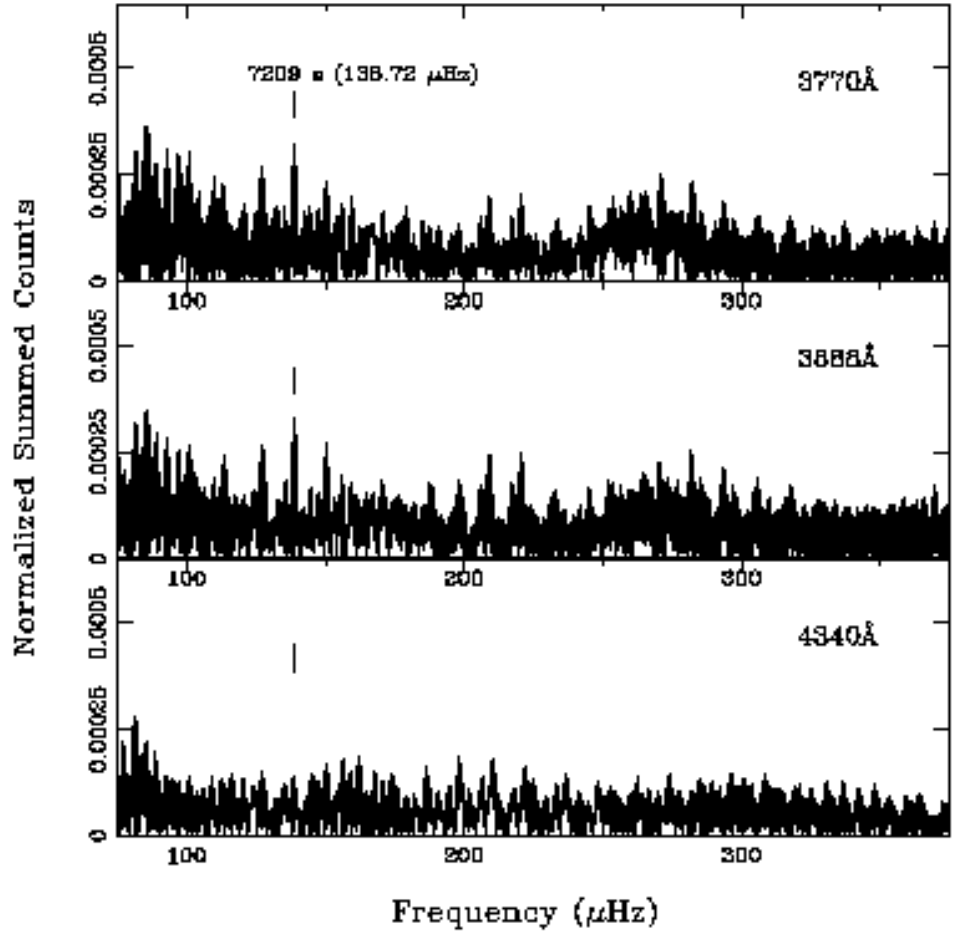}
\figcaption{A comparison of the relative amplitude power spectra (SO
data only) for a blue, an intermediate, and a red Balmer line, after
normalizing each of the extinction-corrected spectra by the reference
amplitude at 4545\AA.  The vertical line corresponds to
7209~s.\label{fig8}}
\end{figure}

\clearpage
\begin{figure}
\epsscale{.80}
\plotone{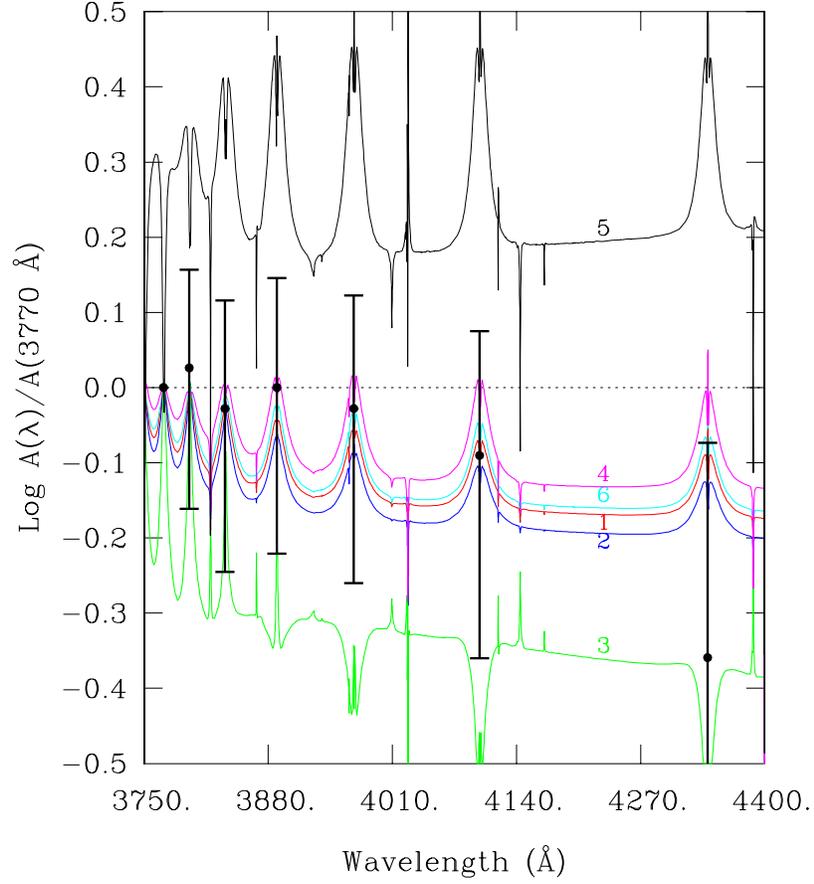}
\figcaption{Logarithm of monochromatic amplitude ratios with respect
to the H11 line core at 3770\AA\ in the spectral range where
measurements (the points with the error bars) have been obtained. This
refers to a specific model of PG 1627+017 and for a g-mode with a
period of 7209 s and degree index $l$ = 1 (red), 2 (blue), 3 (green),
4 (magenta), 5 (black), and 6 (cyan).\label{fig9}}
\end{figure}




\notetoeditor{Figures 3 and 9 are color plots.}

\notetoeditor{We request that Figures 3 through 7 be kept at the
largest practical size possible in the journal, instead of being
reduced to fit in a single column, since the details that they contain
are essential to the understanding of the paper.}


\begin{thebibliography}{}
\bibitem[Allard et al.(1994)]{all94} Allard, F., Wesemael, F.,
    Fontaine, G., Bergeron, P., \& Lamontagne, R. 1994,
    \aj, 107, 1565
\bibitem[Armandroff et al.(1995)]{arman95} Armandroff, T.E., Olszewski, E.W.,
    \& Pryor, C. 1995, \aj, 110, 2131
\bibitem[Baran et al.(2005)]{bar05} Baran, A., Pigulski, A., Kozie\l , D., 
    Og\l oza, W., Silvotti, R., \& Zo\l a, S. 2005, 
    \mnras, 360, 737 
\bibitem[Bill\`eres et al.(2000)]{bill00} Bill\`eres, M., Fontaine, G., 
    Brassard, P., Charpinet, S., Liebert, J., \& Saffer, R. A. 2000,
    \apj, 530, 441
\bibitem[Bill\`eres et al.(2002)]{bill02} Bill\`eres, M., Fontaine, G.,
    Brassard, P., \& Liebert, J.  2002, \apj, 578, 515
\bibitem[Bonanno et al.(2003)]{bon03} Bonanno, A., Catalano, S., Frasca, A., 
    Mignemi, G., \& Patern\`o , L. 2003, \aap, 398, 283
\bibitem[Brassard et al.(2001)]{bras01} Brassard, P.,
    Fontaine, G., Bill\`eres, M., Charpinet, S., Liebert, J., \&
    Saffer, R. A.  2001, \apj, 563, 1013
\bibitem[Charpinet et al.(2003)]{char03} Charpinet, S.,
    Fontaine, G., \& Brassard, P. 2003, in White
    Dwarfs, NATO Science Series, Vol.\ 105, eds.\ D.\ de Martino, R.\
    Silvotti, J.-E.\ Solheim, \& R.\ Kalytis (Dordrecht: Kluwer), 69
\bibitem[Charpinet et al.(2005a)]{cha05a} Charpinet, S.,
    Fontaine, G., Brassard, P., Bill\`eres, M., Chayer, P., \& 
    Green, E.M. 2005a, A\&A, 443, 251
\bibitem[Charpinet et al.(1997)]{char97} Charpinet, S.,
    Fontaine, G., Brassard, P., Chayer, P., Rogers, F.J., Iglesias, C.A.,
    \& Dorman, B. 1997, \apj, 483, L123  
\bibitem[Charpinet et al.(1996)]{char96} Charpinet, S.,
    Fontaine, G., Brassard, P., \& Dorman, B. 1996, 
    \apj, 471, L103  
\bibitem[Charpinet et al.(2000)]{char00} Charpinet, S.,
    Fontaine, G., Brassard, P., \& Dorman, B. 2000, 
    \apjs, 131, 223
\bibitem[Charpinet et al.(2005b)]{cha05b} Charpinet, S.,
    Fontaine, G., Brassard, P., Green, E.M., \& Chayer, P. 2005b,
    \aap, 437, 575
\bibitem[Dreizler et al.(2002)]{drei02} Dreizler, S., Schuh, S. L., 
    Deetjen, J. L., Edelmann, H., \&
    Heber, U. 2002, \aap, 386, 249
\bibitem[Edelmann et al.(2005)]{edel05} Edelmann, H., Heber, U.,
    Altmann, M., Karl, C., \& Lisker, T. 2005, \aap, 442, 1023
\bibitem[Fontaine et al.(2003)]{fon03} Fontaine, G.,
    Brassard, P., Charpinet, S., Green, E.M.,
    Chayer, P., Bill\`eres, M., \& Randall, S.K. 2003, 
    \apj, 597, 518
\bibitem[Green et al.(2003)]{green03} Green, E.M., Fontaine, G.,
    Reed, M.D., Callerame, K., Seitenzahl, I.R., White, B.A., 
    Hyde, E.A., \O stensen, R., Cordes, O., Brassard, P., Falter, S., 
    Jeffery, E.J., Dreizler, S., Schuh, S.L., Giovanni, M., Edelmann, H.,
    Rigby, J., \& Bronowska, A. 2003, \apj, 583, L31
\bibitem[Green et al.(2005)]{green05} Green, E.M., For, B.-Q., \&
    Hyde, E.A. 2005, in 14th European Workshop on White Dwarfs, 
    Vol.\ 334, eds.\ D.\ Koester, \& S.\ Moehler (San Francisco: ASP), 363
\bibitem[Green et al.(2004)]{green04} Green, E.M., For, B.-Q., Hyde, E.A.,
    Seitenzahl, I.R., Callerame, K., White, B.A., Young, C.N., Huff, C.S.,
    Mills, J., \& Steinfadt, J.D.R. 2004, \apss, 291, 267
\bibitem[Green et al.(1997)]{green97} Green, E.M., Liebert, J.W.,
    \& Saffer, R.A. 1997, in Third Conference on Faint Blue Stars, 
    ed.\ A.G.D.\ Philip, J.W.\ Liebert, \& R.A.\ Saffer (Schenectady: 
    L.Davis Press), 417
\bibitem[Heber (1986)]{he86} Heber, U. 1986, \aap, 155, 33
\bibitem[Heber et al.(2003)]{heb03} Heber, U., et al.\ 2003, in White
    Dwarfs, NATO Science Series, Vol.\ 105, eds.\ D.\ de Martino, R.\
    Silvotti, J.-E.\ Solheim, \& R.\ Kalytis (Dordrecht: Kluwer), 105
\bibitem[Horne \& Baliunas(1986)]{hor86} Horne, J.H., \& Baliunas, S.L. 
    1986, \apj, 302, 757
\bibitem[Kilkenny et al.(2002a)]{kil02a} Kilkenny, D. 2002a,
    in IAU Coll. 185, Radial and Nonradial Pulsations
    as Probes of Stellar Physics, ed. C. Aerts, T.R. Bedding, \& J.
    Christensen-Dalsgaard (San Francisco:ASP), 356
\bibitem[Kilkenny et al.(2002b)]{kil02b} Kilkenny, D., Bill\`eres, M., 
    Stobie, R.S., Fontaine, G., Shobbrook, R.R., O'Donoghue, D., 
    Brassard, P., Sullivan, D.J., Burleigh, M.R., \& Barstow, M.A. 
    2002b, MNRAS, 331, 399
\bibitem[Kilkenny et al.(1997)]{kil97} Kilkenny, D., Koen, C.,
     O'Donoghue, D., \& Stobie, R.S. 1997, \mnras, 285, 640
\bibitem[Kilkenny et al.(1999)]{kil99} Kilkenny, D., Koen, C., O'Donoghue, 
    D., van Wyk, F., Larson, K.A., Shobbrook, R., Sullivan, D.J., 
    Burleigh, M.R., Dobbie, P.D., \& Kawaler, S.D. 1999, MNRAS, 303, 525
\bibitem[Kjeldsen (2003)]{kj03} Kjeldsen, H. 2003, \apss, 284, 1
\bibitem[Kjeldsen \& Frandsen (1992)]{kj92}  Kjeldsen, H., \& Frandsen S.
    1992, \pasp, 104, 413
\bibitem[Maxted et al.(2001)]{maxted01} Maxted, P.F.L., Heber, U.,
    Marsh, T.R., \& North, R.C. 2001, \mnras, 326, 1391
\bibitem[Maxted et al.(2002)]{maxted02} Maxted, P.F.L., Marsh, T.R., 
    Heber, U., Morales-Rueda, L., North, R.C., \& Lawson, W.A. 2002, 
    \mnras, 333, 231
\bibitem[Morales-Rueda et al.(2003)]{mo03} Morales-Rueda, L.,
    Maxted, P.F.L., Marsh, T.R., North, R.C., \& Heber, U. 2003,
    \mnras, 338, 752
\bibitem[O'Toole et al.(2002)]{otoole02} O'Toole, S.J., Bedding, T.R., 
    Kjeldsen, H., Dall, T.H., \& Stello, D. 2002, \mnras, 334, 471
\bibitem[O'Toole et al.(2000)]{otoole00} O'Toole, S.J., Bedding, T.R., 
    Kjeldsen, H., Teixeira, T.C., Roberts, G., van Wyk, F., Kilkenny, D.,
    D'Cruz, N., \& Baldry, I.K. 2000, \apj, 537, L53
\bibitem[O'Toole et al.(2004)]{otoole04} O'Toole, S.J., Falter, S., Heber, U.,
    Jeffery, C.S., Dreizler, S., Schuh, S.L., \& the MSST+WET teams 2004,
    \apss, 291, 457.
\bibitem[O'Toole et al.(2003)]{otoole03} O'Toole, S.J., J\o rgensen, M.S.,
    Kjeldsen, H., Bedding, T.R., Dall, T.H., \& Heber, U. 2003, \mnras, 340, 856
\bibitem[Oreiro et al.(2004)]{oreiro04} Oreiro, R., Ulla, A., P\'erez Hern\'andez, F., 
    \O stensen, R., Rodr\'iguez L\'opez, C., \& MacDonald, J. 2004, \aap, 418, 243
\bibitem[MIT(1982)]{mit82} Phelps, F.M., III 1982, MIT Wavelength Tables,
    vol.\ 2 (MIT:Cambridge, Mass.)
\bibitem[Press et al. (1992)]{NR} Press, W.H., Teukolsky, S.A., Vetterling, W.T.,
    \& Flannery, B.P. 1992, Numerical Recipes in Fortran 77, Second Edition,
    (Cambridge University Press:Cambridge, Mass.), p.\ 51
\bibitem[Randall et al.(2005a)]{ran05a} Randall, S.K., Fontaine, G., 
    Brassard, P., \& Bergeron, P. 2005a, \apjs, 161, 456
\bibitem[Randall et al.(2005b)]{ran05b} Randall, S.K., Fontaine, G., 
    Green, E.M., Brassard, P., Kilkenny, D., Crause, L., Terndrup, D.M., 
    Daane, A., Kiss, L.L., Jacob, A.P., Bedding, T.R., For, B.-Q., \& Quirion, 
    P.-O. 2005b, \apj, submitted
\bibitem[Randall et al.(2004)]{ran04} Randall, S., Fontaine, G., 
    Green, E., Kilkenny, D., Crause, L., Cordes, O., O'Toole, S., 
    Kiss, L., For, B.-Q., \& Quirion, P.-O. 2004, Ap\&SS, 291, 465
\bibitem[Randall et al.(2005c)]{ran05c} Randall, S.K., Matthews, J.M.,
    Fontaine, G., Rowe, J., Kuschnig, R., Green, E.M., Brassard, P.,
    Chayer, P., Guenther, D.B., Moffat, A.F.J., Rucinski, S., Sasselow,
    D., Walker, G.A.H., \& Weiss, W.W. 2005c, \apj, 633, 460
\bibitem[Reed et al.(2004)]{reed04} Reed, M.D., 
    Green, E.M., Callerame, K., Seitenzahl, I.R., White, B.A.,
    Hyde, E.A., Giovanni, M.K., \O stensen, R., Bronowska, A., Jeffery, E.J.,
    Cordes, O., Falter, S., Edelmann, H., Dreizler, S., \& Schuh, S.L. 2004,
    \apj, 607, 445
\bibitem[Saffer et al.(1994)]{saf04} Saffer, R.A., Bergeron, P., 
    Koester, D., \& Liebert, J. 1994, \apj, 432, 351
\bibitem[Scargle (1982)]{scar82} Scargle, J.D. 1982, \apj, 263, 835
\bibitem[Silvotti et al.(2002)]{sil02} Silvotti, R., \O stensen, R., Heber, U.,
    Solheim, J.-E., Dreizler, S., \& Altmann, M.  2002, \aap, 383, 239
\bibitem[Solheim et al.(2004)]{solheim04} Solheim, J.-E., \O stensen, R., 
    Silvotti, R., \& Heber, U. 2004, Ap\&SS, 291, 419
\bibitem[Telting \& \O stensen (2004)]{telt04} Telting, J.H. \& \O stensen, R.H.
    2004, \aap, 419, 685
\bibitem[]{} Terman, J.L., Taam, R.E., \& Hernquist, L. 1994, \apj, 422, 729
\bibitem[Thompson et al.(2003)]{thomp03} Thompson, S.E., Clemens, 
    J.C., van Kerkwijk, M.H., \& Koester, D. 2003, \apj, 589, 921
\bibitem{} Unno, W., Osaki, Y., Ando, H., \& Shibahashi, H. 1979,
  Nonradial Oscillations of Stars, Tokyo: Univ. Tokyo Press
\bibitem[van Kerkwijk et al.(2000)]{vankerk00} van Kerkwijk, M.H., 
    Clemens, J.C., \& Wu, Y. 2000, \mnras, 314, 209
\bibitem[Wesemael et al.(1992)]{wese92} Wesemael, F., Fontaine, G.,
    Bergeron, P., Lamontagne, R., \& Green, R.F. 1992 \aj,
    104, 203
\bibitem[Woolf et al.(2002a)]{woolf02a} Woolf, V.M., Jeffery, C.S., \& 
    Pollacco, D.L. 2002a, \mnras, 329, 497
\bibitem[Woolf et al.(2002b)]{woolf02b} Woolf, V.M., Jeffery, C.S., \& 
    Pollacco, D.L. 2002b, \mnras, 332, 34

\end{thebibliography}
\end{document}